\documentclass[journal]{IEEEtran}
\usepackage{graphicx}
\usepackage{amsmath}
\usepackage{multirow}
\usepackage{ragged2e}
\usepackage{makecell}
\usepackage{amssymb}
\usepackage{pifont}
\usepackage{cite}
\usepackage{xcolor}

\ifCLASSINFOpdf
\else
\fi
\begin{document}
%
\title{CI-ICM: Channel Importance-driven Learned Image Coding for Machines}
%
%
%

\author{Yun Zhang~\IEEEmembership{Senior Member,~IEEE}, Junle Liu, Huan Zhang, Zhaoqing Pan~\IEEEmembership{Senior Member,~IEEE}, 
\\ Gangyi Jiang~\IEEEmembership{Senior Member,~IEEE}, 
 and Weisi Lin~\IEEEmembership{Fellow,~IEEE}
\thanks{Yun Zhang and Junle Liu are with the School of Electronics and Communication Engineering, Shenzhen Campus, Sun Yat-Sen University, Shenzhen 518107, China. (Email: zhangyun2@mail.sysu.edu.cn, liujle@mail2.sysu.edu.cn).}
\thanks{Huan Zhang is with the School of Information Engineering, Guangdong University of Technology, Guangzhou 510006, China. (Email: huanzhang2021@gdut.edu.cn)}
\thanks{Zhaoqing Pan is with the School of Electrical and Information Engineering, Tianjin University, Tianjin 300072, China. (Email: pan\_zhaoqing@hotmail.com)}
\thanks{Gangyi Jiang is with the Faculty of Electrical Engineering and Computer Science, Ningbo University, Ningbo 315211, China. (Email: jianggangyi@nbu.edu.cn)}
\thanks{Weisi Lin is with the College of Computing and Data Science, Nanyang Technological University, Singapore
(e-mail: wslin@ntu.edu.sg).}
}

%
%

\markboth{IEEE Transactions on Circuits and Systems for Video Technology,~Vol.xx, No.xx, 2025}%
{Shell \MakeLowercase{\textit{et al.}}: Bare Demo of IEEEtran.cls for IEEE Journals}
%



\maketitle

\begin{abstract}
\justifying{
Traditional human vision-centric image compression methods are suboptimal for machine vision centric compression due to different visual properties and feature characteristics. To address this problem, we propose a Channel Importance-driven learned Image Coding for Machines (CI-ICM), aiming to maximize the performance of machine vision tasks at a given bitrate constraint. First, we propose a Channel Importance Generation (CIG) module to quantify channel importance in machine vision and develop a channel order loss to rank channels in descending order. Second, to properly allocate bitrate among feature channels, we propose a Feature Channel Grouping and Scaling (FCGS) module that non-uniformly groups the feature channels based on their importance and adjusts the dynamic range of each group. Based on FCGS, we further propose a Channel Importance-based Context (CI-CTX) module to allocate bits among feature groups and to preserve higher fidelity in critical channels. Third, to adapt to multiple machine tasks, we propose a Task-Specific Channel Adaptation (TSCA) module to adaptively enhance features for multiple downstream machine tasks. Experimental results on the COCO2017 dataset show that the proposed CI-ICM achieves BD-mAP@50:95 gains of 16.25$\%$ in object detection and 13.72$\%$ in instance segmentation over the established baseline codec. Ablation studies validate the effectiveness of each contribution, and computation complexity analysis reveals the practicability of the CI-ICM. This work establishes feature channel optimization for machine vision-centric compression, bridging the gap between image coding and machine perception.}
\end{abstract}


%
\IEEEpeerreviewmaketitle

\section{Introduction}
\label{sec:introduction}

\IEEEPARstart{I}{n} recent years, intelligent applications such as smart cities \cite{digitalretina} and Internet of Things (IoT) systems \cite{iot, iot2} have significantly boosted demands for efficient visual data transmission between edge devices and centralized cloud intelligent analytic models, which require advanced visual coding technologies. Currently, learned codecs have advanced rapidly, with state-of-the-art codecs that exceed traditional standards such as JPEG and Versatile Video Coding (VVC) \cite{vvc} in compression efficiency. However, these codecs prioritize minimizing perceptual distortions for human viewers, which have proven suboptimal for machine vision, and machine-specific requirements are critical for future codecs \cite{HNRISC}. To address this urgent demand, the Moving Picture Experts Group (MPEG) has launched standardization activities for Image/Video Coding for Machines (ICM/VCM) \cite{vcm}, which prioritizes machine vision-oriented compression. As highlighted in \cite{vcmaparadigm, TDVCM}, ICM represents a shift in visual data processing for the next generation of intelligent systems and multimedia applications. The key to developing an efficient ICM codec lies in exploiting the redundancies in the pixel domain and the feature domain. Thus, ICM is categorized into pixel-based \cite{JinJND2022, TAQNforJPEG, analysisRD2021, SaliencyBA, zhangyun2024, towardsSIC2021, shindo2024image, zhang2023rethinking} and feature-based schemes \cite{endtoend2021wang, ICMcontent2021, LeICM2021, end-to-end, matsubara2023sc, liu2025multiscale, endtoendICM2021, Rate-DistortionTheory2025, improvingmvtincompressdomain, scalableicm_choi, Choi2021LatentSpaceSF, zhang2024, latentSICM, mutualFCMV2023, SSICMfeature, Cui2025, Chen2023TransTICTT, Li2024ImageCF}. 

In pixel-based schemes, Jin \emph{et al.} \cite{JinJND2022} proposed a just noticeable difference model for image classification that analyzed the spatial preferences of images for machine vision, reflecting the core motivation of pixel-based schemes. Some pixel-based schemes are based on traditional codecs such as JPEG and VVC to improve the ICM coding efficiency. Choi \emph{et al.} \cite{TAQNforJPEG} presented a neural network to predict task-specific quantization maps for JPEG. Huang \emph{et al.} \cite{analysisRD2021} proposed a region-of-interest-based bit allocation algorithm, which evaluates the importance of each coding tree unit and allocates more bits for those of high importance. Li \emph{et al.} \cite{SaliencyBA} proposed a saliency segmentation-oriented bit allocation for learned image compression, where pixel-wise salient regions are assigned with more bits to maintain high segmentation accuracy at a given bit rate. Zhang \emph{et al.} \cite{zhangyun2024} developed Just Recognizable Distortion (JRD) models to predict the visibility threshold of machine vision, which was applied to enhance the VVC based ICM. These approaches are based on traditional human-oriented codecs, which exhibit limited compatibility with machine vision-oriented coding. 

With the rapid development of deep learning-based codecs, other pixel-based methods rely on these learned codecs for better coding efficiency. Yang \emph{et al.} \cite{towardsSIC2021} proposed a scalable face image coding scheme for facial detection, which separates facial contour and color information, and prioritizes contour during compression. Shindo \emph{et al.} \cite{shindo2024image} proposed an ICM codec that focuses on edge information of object parts in an image. Taking this further, Zhang \emph{et al.} \cite{zhang2023rethinking} proposed a scalable cross-modality compression scheme, which achieves more granular decomposition by separating images into semantic, structural, and signal components. These pixel-based optimizations involve the decomposition of spatial information into distinct, task-specific components, which enables simplicity and generalization ability in design. However, these methods do not fully take advantage of the potent semantic representation power in the feature space, resulting in suboptimal ICM performance.

In contrast to pixel-based approaches, feature-based encoding schemes for ICM focus on optimizing latent representation within learned codecs to improve coding performance. Some deploy learned image codecs within the ICM process and train them. Wang \emph{et al.} \cite{endtoend2021wang} presented a joint training paradigm by introducing a Lagrange multiplier in training loss to navigate the trade-off between reconstruction quality, task performance, and bitrate. Le \emph{et al.} \cite{ICMcontent2021}{\cite{LeICM2021}} proposed an ICM codec trained by knowledge distillation, which uses a pre-trained task network to guide the reconstructed features. {These methods concatenate end-to-end learned codecs with task models for joint training. Since these learned codecs are designed for human vision rather than machine vision, some approaches further optimize the architecture of the codec for machine vision to improve ICM performance.} In other joint training schemes, codecs are deployed within machine vision models to compress the intermediate features of task models \cite{end-to-end}. Matsubara \emph{et al.} \cite{matsubara2023sc} proposed a supervised compression approach for split computing, which transmits intermediate features with lower bitrates. Liu \emph{et al.} \cite{liu2025multiscale} proposed a bit allocation method for end-to-end feature compression by exploiting the importance of multiscale features. Chamain \emph{et al.} \cite{endtoendICM2021} conducted an analysis of different joint training configurations, showing that we can selectively train the encoder, decoder, or task network and achieve rate-accuracy improvements over a pretrained codec. {Harell \emph{et al.} \cite{Rate-DistortionTheory2025} presented a formulation of rate-distortion theory for the joint training paradigm and demonstrated improved coding performance with task-specific approaches. Their analysis demonstrates that transmitting task results maximizes compression efficiency. However, this approach lacks the flexibility required for real-world deployment. Transmitting image features remains essential to accommodate varying and evolving downstream tasks.} In summary, these joint training strategies rely on existing compression architectures and align distorted features with their original counterparts, leaving significant room for improvement through fine-grained optimization in the feature representation.

Building upon joint training ICM, recent studies investigated feature representation optimizations in learned image codecs for better coding efficiency. Much effort has been devoted to separating features into task-crucial and non-crucial subsets. Liu \emph{et al.} \cite{improvingmvtincompressdomain} proposed a gating mechanism in the encoder to select features targeting different tasks. Choi \emph{et al.} \cite{scalableicm_choi, Choi2021LatentSpaceSF} proposed scalable coding methods by directly separating machine-oriented and human-oriented features from the encoder, and only transmitting machine-oriented features in the ICM scenario. These methods separate features along the channel dimension, but do not differentiate the importance of each channel for machine vision. This overlook leads to an ambiguous understanding of the importance distribution, which does not provide clear guidance for optimization. Therefore, some approaches attempt to further analyze and exploit the feature characteristics. Zhang \emph{et al.} \cite{zhang2024} presented a deep image codec for hybrid human-machine vision tasks and designed a scalable entropy model to remove redundancy between semantic and visual features. Özyılkan \emph{et al.} \cite{latentSICM} presented an ICM codec that uses feature division methods to extract and separate semantic features. In \cite{mutualFCMV2023}, the proposed ICM method only compresses part of the feature pyramid to save bitrate and reconstructs the complete feature based on these partial features. Yan \emph{et al.} \cite{SSICMfeature} proposed a scalable coding scheme, which hierarchically extracts semantic features from fine to coarse and transmits selected features according to specific tasks. However, they rely heavily on the primary task loss to guide feature separation. It is task-specific and has a high risk of misclassification of other tasks without explicit constraints. The generalization to multi-tasks could be improved. 
The principle of feature separation is ambiguous and implicit, which may cause misclassifications leading to information redundancies and task precision loss. 

Besides feature separation approaches, several feature-based methods adapt the extracted features to ICM. Cui \emph{et al.} \cite{Cui2025} employed an implicit semantic module to adapt the extraction of machine-oriented features. Chen \emph{et al.} \cite{Chen2023TransTICTT} proposed an adaptation method from human-centric learned image compression to machine-centric ICM by leveraging a prompt generator, called TransTIC. It assigns instance-specific prompts to the encoder and task-specific prompts to the decoder. Li \emph{et al.} \cite{Li2024ImageCF} proposed a modulation adapter for ICM, named AdaptICMH, which uses a spatial modulation adapter to remove non-semantic redundancy and a frequency modulation adapter to amplify task-relevant components while suppressing the irrelevant ones. These feature-based adaptation methods take advantage of a learned image compression framework and are adapted to ICM for high coding efficiency by further exploiting machine vision redundancies. However, these adaptations mainly fine-tuned the extracted features and did not alter the feature structure, leading to limited improvements. The importance of feature channels is not well considered and could be further exploited.

To overcome limitations on previous ICMs and address multi-task adaptation, we propose a Channel Importance-driven learned Image Coding for Machines (CI-ICM), which consists of Channel Importance Generation (CIG) for importance analysis, Feature Channel Grouping and Scaling (FCGS) and Channel Importance-based Context (CI-CTX) for bit allocation, and Task-Specific Channel Adaptation (TSCA) for task adaptation. The main contributions are as follows:
\begin{itemize}
\item We propose a novel CI-ICM framework to improve the compression efficiency for machine vision by exploiting the importance of feature channels. Compared with ELIC, it achieves an average of 16.25$\%$ BD-mAP@50:95 gains in object detection and 13.72$\%$ BD-mAP@50:95 gains in instance segmentation, which are superior to the state-of-the-art TransTIC and AdaptICMH schemes.
\item We propose a CIG module to explicitly analyze the importance of feature channels. Complemented by a novel channel order loss, CI-ICM extracts the ordered feature representation and efficiently separates features into subsets representing distinct levels of importance. 
\item We propose two collaborative modules, FCGS and CI-CTX, to determine the scaling of features and allocate the bitrate based on the importance of feature channels. 
\item We design a TSCA module to improve feature representation and adapt multiple machine tasks by taking advantage of the importance of task-specific channels. 
\end{itemize}


The paper is organized as follows: Section~\ref{motivation} presents the motivation and analysis. Section~\ref{proposed method} presents the proposed CI-ICM framework, key modules, loss functions, and training procedure. Section~\ref{exp results and analysis} presents the experimental results for different tasks, ablation studies, and complexity analysis. Finally, Section~\ref{conclusion} draws the conclusion.

 

\section{Motivations and Analysis}
\label{motivation}
Feature channels are of different importance to machine vision. To validate this assumption, we manually add distortion and analyze the different impacts of feature channels on machine vision. A practical ICM scenario involves different machine vision tasks, which require a highly versatile codec. We set object detection as the primary machine vision task and instance segmentation as the secondary task for subsequent evaluations. The machine vision task is conducted using a Faster R-CNN~\cite{fasterrcnn} model, with performance benchmarked on the COCO2017 dataset. Performance was quantified using the mean Average Precision at Intersection over Union (IoU) thresholds ranging from 50$\%$ to 95$\%$, which is denoted mAP@50:95. The features we assess are extracted from the established ICM codec~\cite{Chen2023TransTICTT}. To analyze the importance of each feature channel, we conducted four types of coding experiments by adding distortion to feature channels, which zero out or add random noise to an individual feature channel and one group of feature channels. {Specifically, 192 features are evenly divided into 8 groups along the channel dimension, which are 24 channels for each group.}


\begin{figure}
\begin{minipage}[b]{0.49\linewidth}
  \centering
  \centerline{\includegraphics[width=\linewidth]{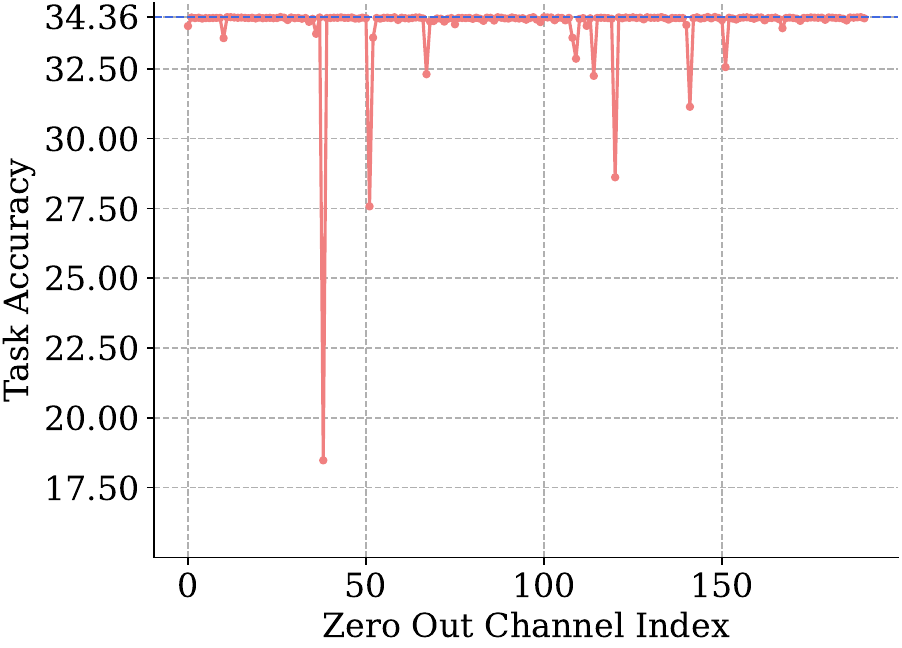}}
  \centerline{(a)}\medskip
\end{minipage}
\hfill
\begin{minipage}[b]{0.49\linewidth}
  \centering
  \centerline{\includegraphics[width=\linewidth]{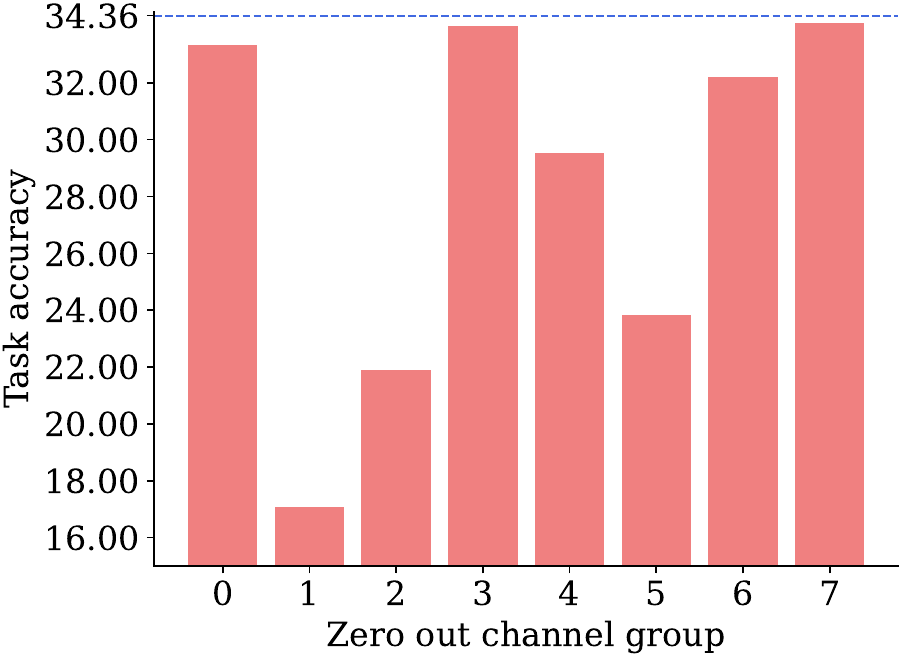}}
  \centerline{(b)}\medskip
\end{minipage}

\begin{minipage}[b]{0.49\linewidth}
  \centering
  \centerline{\includegraphics[width=\linewidth]{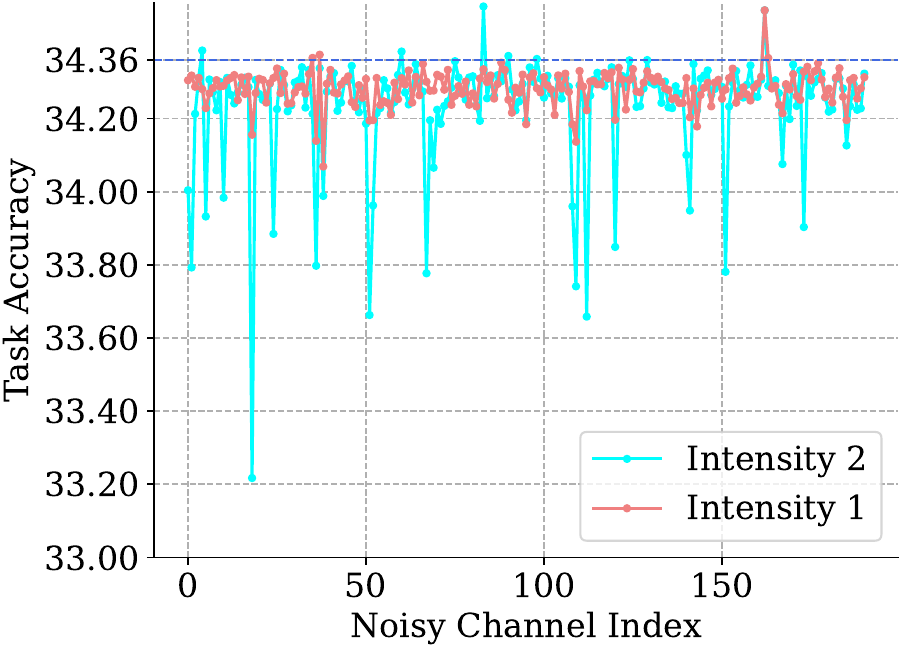}}
  \centerline{(c)}\medskip
\end{minipage}
\hfill
\begin{minipage}[b]{0.49\linewidth}
  \centering
  \centerline{\includegraphics[width=\linewidth]{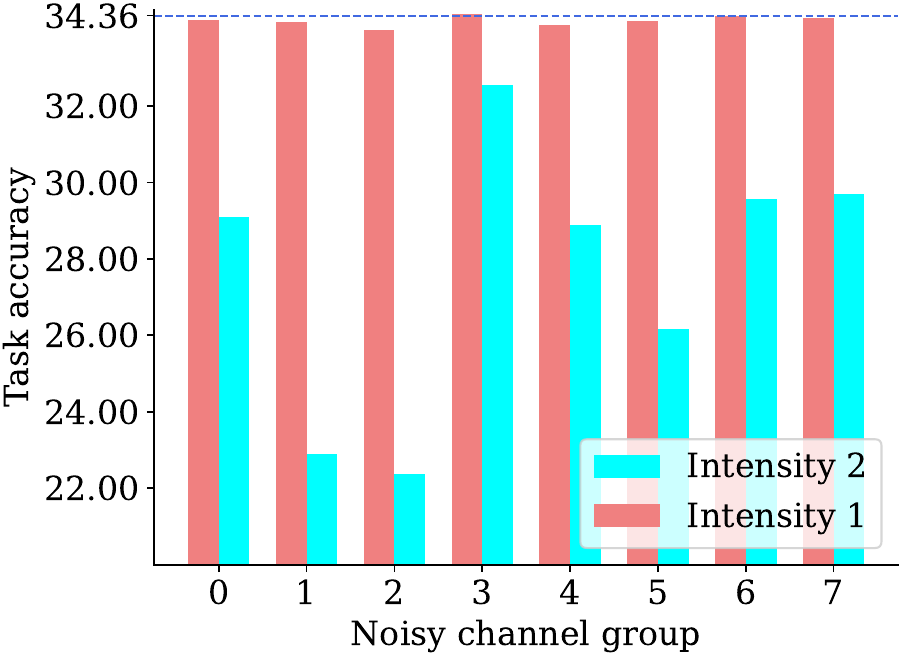}}
  \centerline{(d)}\medskip
\end{minipage}
\caption{Feature channel importance analysis by adding distortions, where mAP@50:95 was used as precision metric for object detection. (a) zero out for individual channel, (b) zero out for a group of channels, (c) distortions added to individual channel, where “Intensity 1” and “Intensity 2” indicated the magnitudes of the added random noise ranges $[-0.5,0.5]$ and $[-1,1]$, respectively, (d) distortions added to a group of channels.}
\label{fig:motivation}
\end{figure}

\begin{figure*}[!ht]
\begin{minipage}[b]{0.68\linewidth}
  \centering
  \centerline{\includegraphics[width=\linewidth]{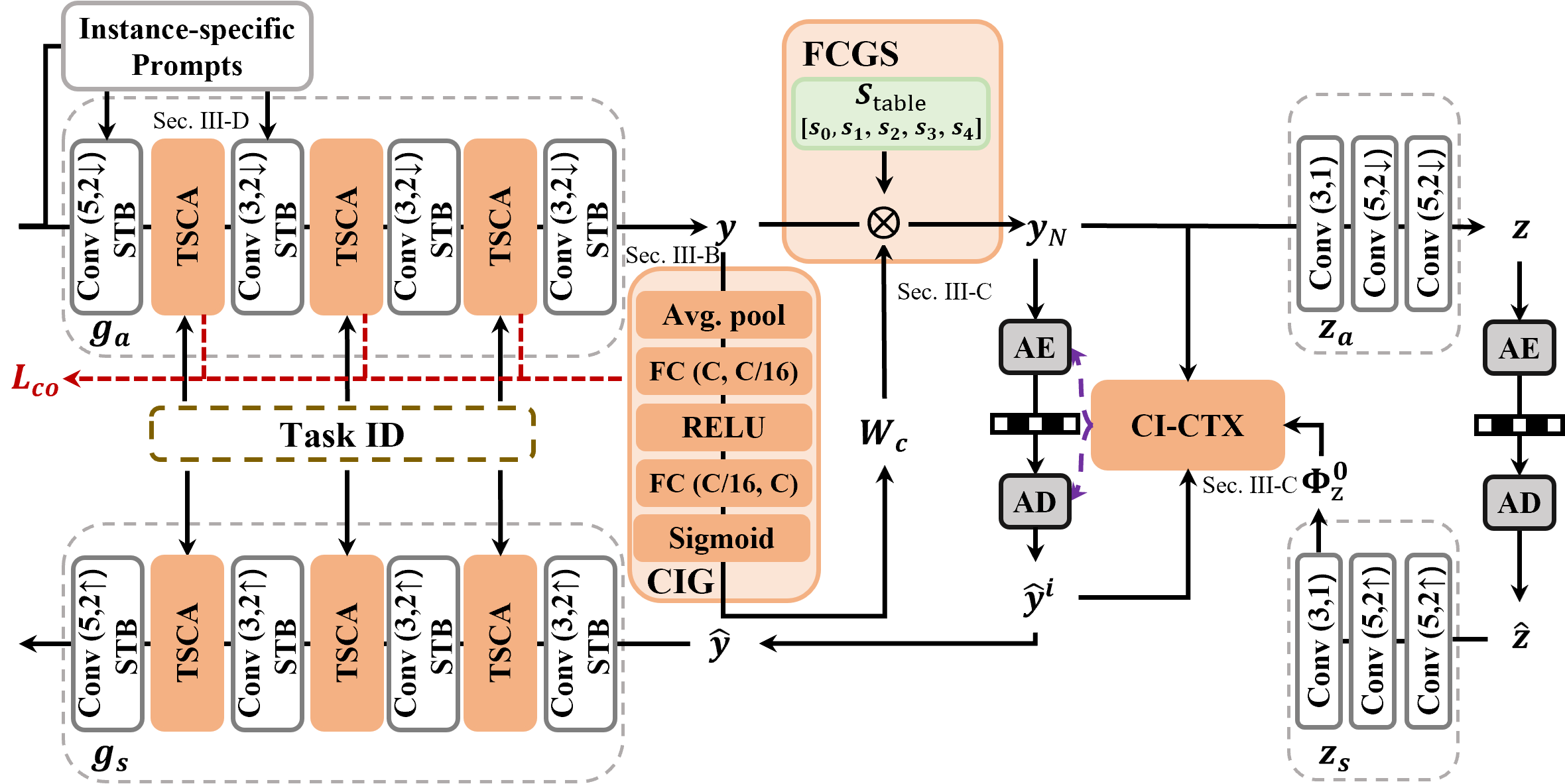}}
  \centerline{(a)}\medskip
\end{minipage}
\hfill
\begin{minipage}[b]{0.31\linewidth}
  \centering
  \centerline{\includegraphics[width=\linewidth]{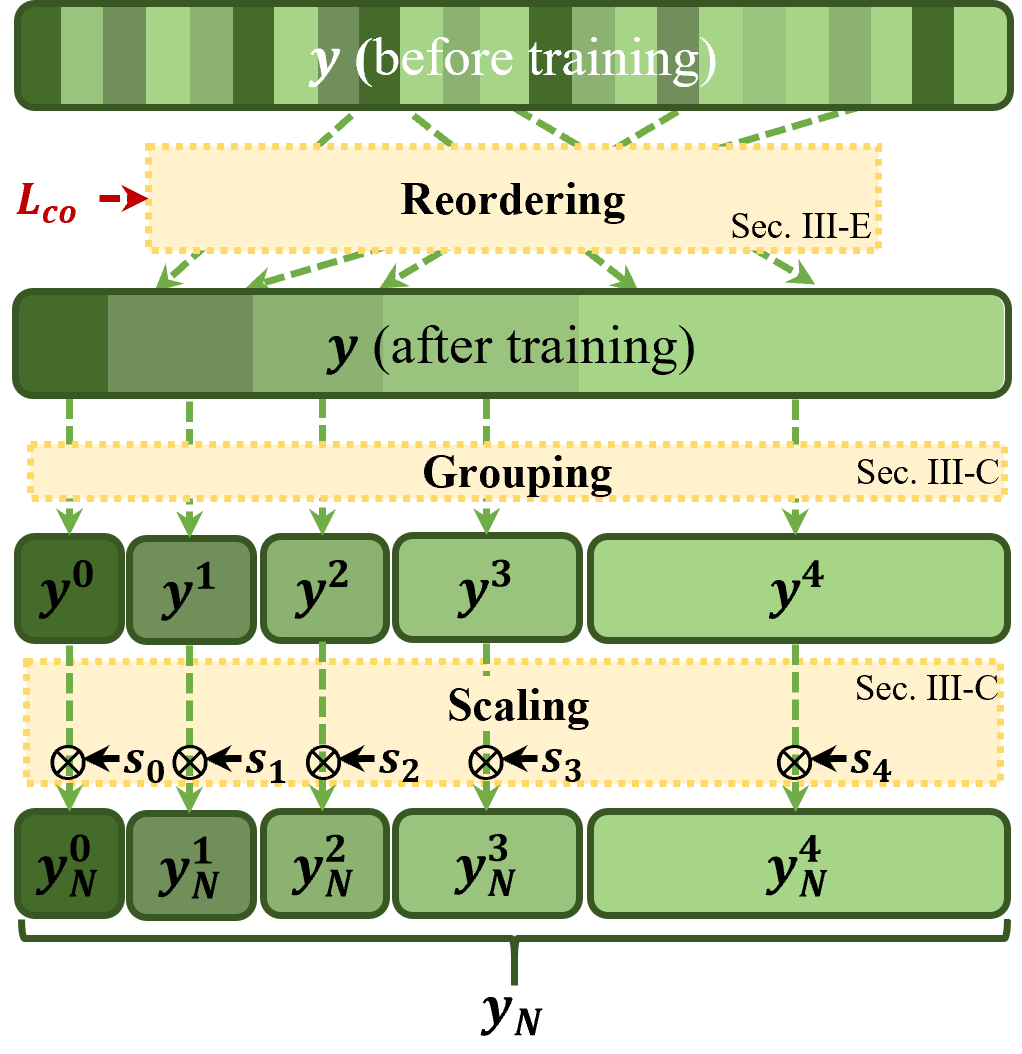}}
  \centerline{(b)}\medskip
\end{minipage}
\caption{Framework of the proposed CI-ICM. (a) Network architecture of the proposed CI-ICM, (b) The reordering, grouping, and scaling feature flow of $\mathbf{y}$, where darker green $\mathbf{y}$ blocks are features with higher importance for machine vision.}
\label{fig:pipeline}
\end{figure*}

Fig.1 shows the experimental analysis on the impacts of distortions in feature channels, where the x-axis is the feature or feature group index and the y-axis is the task accuracy measured with mAP@50:95. We have the following observations: 1) when zeroing out one of the individual features, the precision degradation varies significantly. Similarly, when zeroing out one group of features, the precision degrades significantly for some groups such as \#1, \#2, and \#5. The precision remains almost the same for groups \#0, \#3, and \#7. It indicates that the channel importance varies in zero out distortion. 2) When adding distortion with different magnitudes, e.g., ranging [-0.5,0.5] and [-1,1], to an individual channel and one of the groups, the task precision remains almost the same for Intensity 1 and degrades significantly for Intensity 2. It indicates that the distortion impacts not only vary with feature channels, but also vary with the intensities of distortion. These feature importances can be exploited to improve image coding efficiency.


\section{The Proposed CI-ICM}
\label{proposed method}
In this section, we present the proposed CI-ICM and key modules, including CIG, FCGS, CI-CTX, and TSCA. In addition, loss functions and the training procedure are presented.

\subsection{Framework of the Proposed CI-ICM}
\label{pipeline}
We propose CI-ICM to take advantage of feature channel importance to improve ICM efficiency. Fig.~\ref{fig:pipeline}(a) shows the architecture of the proposed CI-ICM framework. It includes encoder $g_a$ to extract latent features, decoder $g_s$ for image reconstruction, hyper-encoder $z_a$ and hyper-decoder $z_s$ for the extraction and reconstruction of side information, arithmetic encoder AE and decoder AD for entropy coding. In addition, we propose new coding modules to enhance the ICM coding efficiency, including CIG to analyze channel importance, FCGS and CI-CTX for bitrate allocation, TSCA for task adaptation and high efficiency, which are highlighted.

Initially, the encoder $g_a$ processes an input image to obtain the latent space representation $\mathbf{y}$. The CIG evaluates $\mathbf{y}$ to produce a coefficient vector $\mathbf{W}_{c}$, quantifying and predicting the importance of each feature channel, which was then used to guide feature representation and coding. 
The latent space representation $\mathbf{y}$ is scaled and processed with FCGS to produce $\mathbf{y}_N$. The FCGS module reorders the features, separates the features into uneven channel groups, and adjusts the dynamic range of different groups, which can be exploited in entropy coding for high efficiency. The scaled representation $\mathbf{y}_N$ is then fed into CI-CTX to model the entropy probability of the unevenly grouped $\mathbf{y}_N$ sequentially, which improves the entropy coding and decoding. Feature channels of higher importance are grouped with a small group size, which are encoded earlier to allow the CI-CTX to allocate more bits. The feature channels with lower importance are grouped with a larger group size and encoded later, which will be allocated with fewer bits to improve the compression ratio. Finally, the reconstructed $\hat{\mathbf{y}}$ is input to decoder $g_s$ to reconstruct an output image, which is input to the machine vision task models. To adapt to multiple different machine tasks, we proposed TSCA modules in both the proposed encoder $g_a$ and decoder $g_s$, which can dynamically adjust channel-wise enhancement according to the target task. 

In this work, we use TransTIC \cite{Chen2023TransTICTT} as the base ICM backbone and use the instance-specific prompt tuning mechanism to ensure robust spatial adaptation for machine vision. The hyper-prior encoder $z_a$ and the decoder $z_s$ from \cite{He2022ELICEL} are used for effective extraction and side information compression.

\subsection{Proposed CIG Module}
\label{cig}
To differentiate the priorities of feature channels based on their contribution to machine vision tasks, we propose the CIG module to quantify the importance of each channel within the latent representation $\mathbf{y}$. 
Given the CIG module $\mathcal{M}_{c}(\cdot)$, this is achieved by generating a importance vector $\mathbf{W}_{c}$, which then modulates the features $\mathbf{y}$ to produce $\mathbf{y}_{out}$. This process is presented as
\begin{equation}
\begin{cases}
    & \mathbf{W}_{c} = \mathcal{M}_{c}(\mathbf{y}) \\
    & \mathbf{y}_{out} = \mathbf{y} \cdot \mathbf{W}_{c}
\end{cases}.
\end{equation}
Fig.~\ref{fig:pipeline}(a) also shows the working flow of the CIG module, which includes an average pooling layer followed by a network consisting of two fully connected layers with a ReLU activation function between them, and a final sigmoid activation function. 
According to~\cite{senet}, these weights $\mathbf{W}_{c}$ quantify channel importance, with critical channels assigned larger weights.

\begin{figure}
\begin{minipage}[b]{0.49\linewidth}
  \centering
  \centerline{\includegraphics[width=\linewidth]{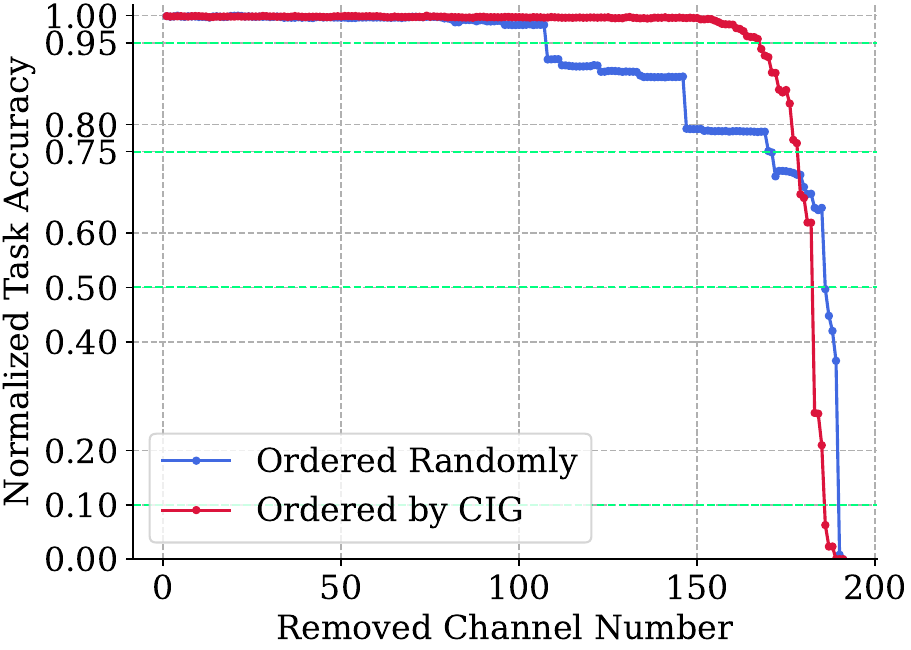}}
  \centerline{(a)}\medskip
\end{minipage}
\hfill
\begin{minipage}[b]{0.49\linewidth}
  \centering
  \centerline{\includegraphics[width=\linewidth]{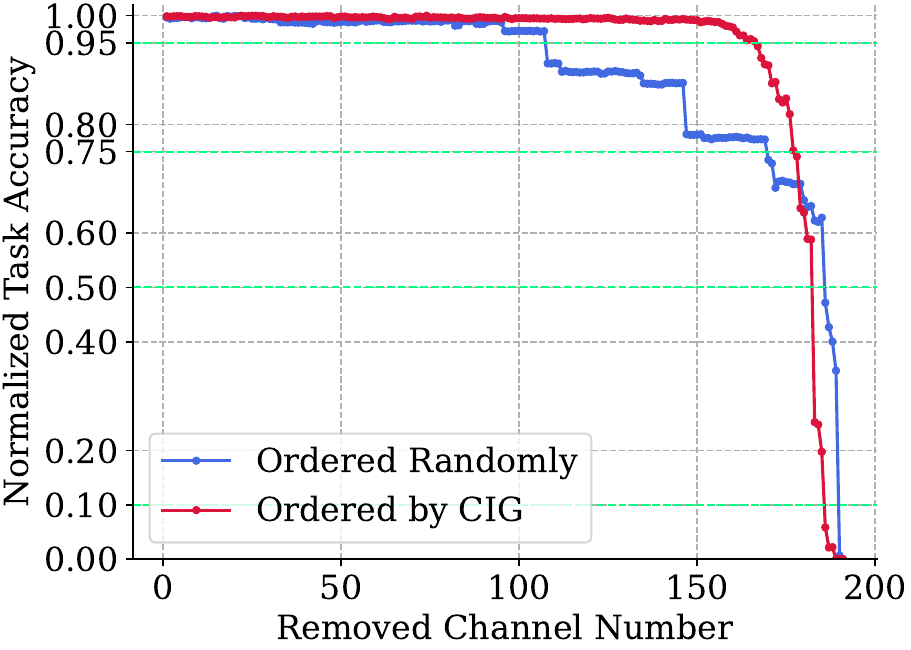}}
  \centerline{(b)}\medskip
\end{minipage}
\caption{Relationship between the number of removed channels and the normalized object detection task accuracy. (a) Normalized mAP@50:95, (b) Normalized mAP@75.}
\label{fig:set0}
\end{figure}

To train and validate the effectiveness of the proposed CIG module, we performed a feature channel removal experiment with different ordering settings.
We adopted the baseline TransTIC coding framework~\cite{Chen2023TransTICTT} and integrated only the CIG module to process the latent feature $\mathbf{y}$. We trained the CIG while keeping all other parameters frozen to obtain the channel importance weights, $\mathbf{W}_{c}$. Then, we progressively removed the feature channels from lowest to highest importance weights, labeled “Ordered by CIG”. For comparison, the channels were randomly ordered and then also removed in a random order, labeled “Ordered Randomly”. Then, the removed and ordered features were input to TransTIC for encoding and decoding. The reconstructed images were input to task model Faster R-CNN for object detection. Mean Average Precision at IoU thresholds from 50$\%$ to 95$\%$ (mAP@50:95) and at 75$\%$ (mAP@75) were used for object detection performance evaluation of coded images. Then, the mAP values are normalized to [0,1], which are denoted as normalized mAP@50:95 and mAP@75. The coding experiment to analyze the impacts of feature channel ordering included all images from COCO2017 validation dataset, which consists of 5,000 images, 91 categories, and 5 captions per image.

Fig.~\ref{fig:set0} shows the relationship between object task precision and the number of removed feature channels. There are 192 channels in total for the feature representation $\mathbf{y}$. We can observe that for randomly ordered features, the task precision starts to degrade when the number of removed features increases to 110. More degradations are caused when more feature channels are removed. For the features ordered based on feature importance weights $\mathbf{W}_{c}$ from CIG, the object detection precision degrades when the number of removed feature channels increases to 160. It indicates that, based on the weights $\mathbf{W}_{c}$, more features of higher importance can be reserved in the feature removal experiment, which validates the effectiveness of CIG and its generated importance weights.

Based on these weights $\mathbf{W}_{c}$, we propose a channel order loss $L_{CO}$ to guide the extraction process of the feature representation module $g_a$. In this case, the output feature $\mathbf{y}$ from retrained $g_a$ will be reordered in descending order as $\mathbf{y}_{out}$  based on their contributions to the target machine task, i.e., $\mathbf{W}_{c}$. Details of the loss $L_{CO}$, module training, and validation will be presented in Section \ref{COL}.


\subsection{Proposed FCGS and CI-CTX Modules}
\label{sm}

\subsubsection{Architectures}
\label{sm_1}
Based on these weights from the CIG and reordered features, we subsequently propose FCGS to further process the feature representation for high-efficiency ICM. Fig.~\ref{fig:pipeline}(b) shows the data flow of processing the reordered features $\mathbf{y}$, which are grouped and scaled based on the channel importance. First, since the features are not equally important, shown as Fig.~\ref{fig:set0}, the features are divided into $n$ uneven groups, denoted as $\{\mathbf{y}^{i}\}$, $i\in[0,n-1]$. The key principle of this division is that features of higher importance are put into smaller groups, and vice versa. Then, the grouped features are scaled by using a scaling table, $\{{s}_{i}\}$, $i\in[0,n-1]$, generating $\mathbf{y}_N^{i}$, correspondingly as 
\begin{equation}
    \mathbf{y}_N^{i} = \mathbf{y}^{i}/s_i.
\end{equation}
Smaller scaling factors are assigned to feature groups with higher importance to maintain high fidelity. On the other hand, larger scaling factors will be given to feature groups with lower importance, which targets to lower the coding bit rate. As $\mathbf{y}^{0}$ is the most critical group, $s_0$ is set to $1$. The rest $s_i$ is larger than 1, which will be determined experimentally.

\begin{figure}
\centering
\centerline{\includegraphics[width=\linewidth]{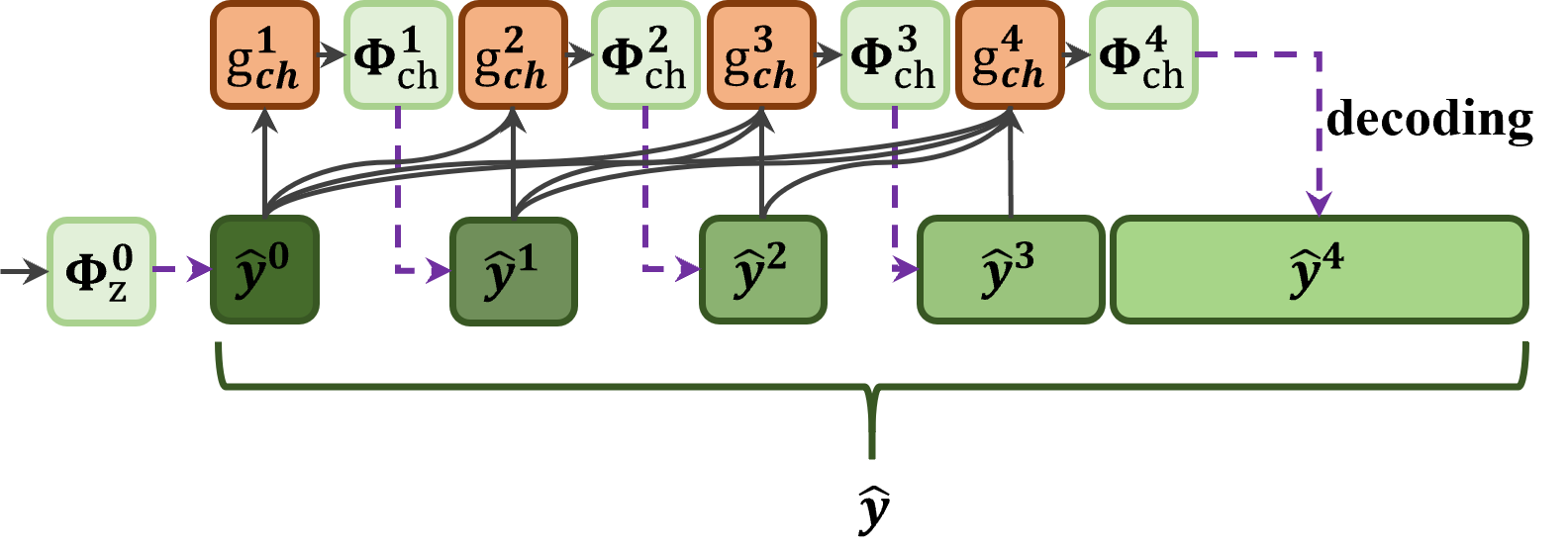}}
\caption{Decoding process of the CI-CTX module.}
\label{fig:ci-ctx}
\end{figure}

Inspired by \cite{He2022ELICEL}, we propose CI-CTX to properly allocate bits between different feature groups, leveraging the importance of the feature channel, which aims to further improve coding efficiency. As the FCGS divided the feature channels into a number of uneven subsets, the previous and more important features can be referenced as prior in decoding subsequent feature channels. Fig.~\ref{fig:ci-ctx} shows the decoding process, where the uneven groups $[\hat{\mathbf{y}}^{0}, \hat{\mathbf{y}}^{1}, ..., \hat{\mathbf{y}}^{n-1}]$ are decoded sequentially. $\hat{\mathbf{y}}^0$ is decoded based on hyper-prior information $\mathbf{\Phi}_{\text{z}}^0$. Subsequent groups are decoded in sequence, and each group relies on prior information derived from previously decoded groups as
\begin{equation}
\begin{cases}
    & \mathbf{\Phi}_{\text{z}}^0 = z_s(AD(AE(z_a(\mathbf{y}_N)))), \\
    & \mathbf{\Phi}_{\text{ch}}^i = g_{\text{ch}}^i (\hat{\mathbf{y}}^{0}, \hat{\mathbf{y}}^{1}, ..., \hat{\mathbf{y}}^{i-1}),
\end{cases}
\end{equation}
where $\mathbf{\Phi}_{\text{z}}^0$ denotes the hyper-prior parameters for decoding $\hat{\mathbf{y}}^0$, $z_a(\cdot)$ and $z_s(\cdot)$ denote the hyper-encoder and hyper-decoder, $AE(\cdot)$ and $AD(\cdot)$ denote the arithmetic encoding and decoding, $\mathbf{\Phi}_{\text{ch}}^i$ is the channel-wise context prior parameters of the $i$-th group, and $g_{\text{ch}}^i(\cdot)$ represents the prior extraction network of the $i$-th channel.

As FCGS allocates more important channels to smaller groups, in the CT-CTX coding process, the former coded important groups contain a smaller number of feature channels, where more bits will be allocated to each feature channel. In addition, their priors are more representative and are referenced in subsequent coding by exploiting the channel correlation. Based on the ordering, uneven grouping, and scaling processes of FCGS and CI-CTX, more coding bits are allocated to features of higher importance, leading to better coding efficiency. Optimal parameter determination for FCGS and CI-CTX will be presented in the next subsection. Moreover, while coding each feature group in CI-CTX, we applied the spatial context module of the parallel checkerboard method~\cite{Jiang2022MLICME} to further improve the coding efficiency. 

\subsubsection{Optimal Parameter Determination}
\label{sm_2}
To properly allocate the bitrate in FCGS and CI-CTX, three key parameters must be determined, namely the number of groups $n$, the sizes of each feature group $\{l_i\}$, and the scaling factors for each group $\{s_i\}$. 
A large $n$ will increase the computational complexity without additional coding gains, while the channel context cannot be effectively exploited with a small $n$. In \cite{He2022ELICEL}, all feature channels are divided into five groups to achieve a good trade-off between coding gain and complexity. So, in this work, $n$ is set as 5 for fair comparisons, indicating that all 192 feature channels in ${\mathbf{y}}$ are divided into five groups.

Then, the size of each feature channel group $\{l_i\}$ shall be determined. Based on the feature importance analysis in Fig. \ref{fig:set0}, we found that in a total of 192 reordered feature channels, the task accuracy almost maintains the same when about 160 features are removed, which implies these 160 channels of features are with less importance. Then, the task accuracy degrades more and more sharply while removing the remaining 32 feature channels, indicating that they are of high importance. Therefore, 160 channels of features are grouped into one set with the least importance, corresponding to $l_{4}=160$. Based on \cite{GOSWAMI2024111921} and the principle that high importance features are set to a small group\cite{He2022ELICEL}, the 32 feature channels are divided into four groups and the size of each group,  corresponding to $\{l_{i}\}=[4,4,8,16]$, $i\in\{0,1,2,3\}$.
The grouping strategy's adaptability is evidenced by the fact that alternative threshold selections can yield similar performance, as it is designed around the task-specific information capacity of the features. The channel order loss ensures that features are ranked by importance, facilitating their division into ordered groups of predetermined size. This structure prioritizes crucial channels by placing them in smaller groups, thereby preserving higher compression quality.

\begin{figure}
\begin{minipage}[b]{0.49\linewidth}
  \centering
  \centerline{\includegraphics[width=\linewidth]{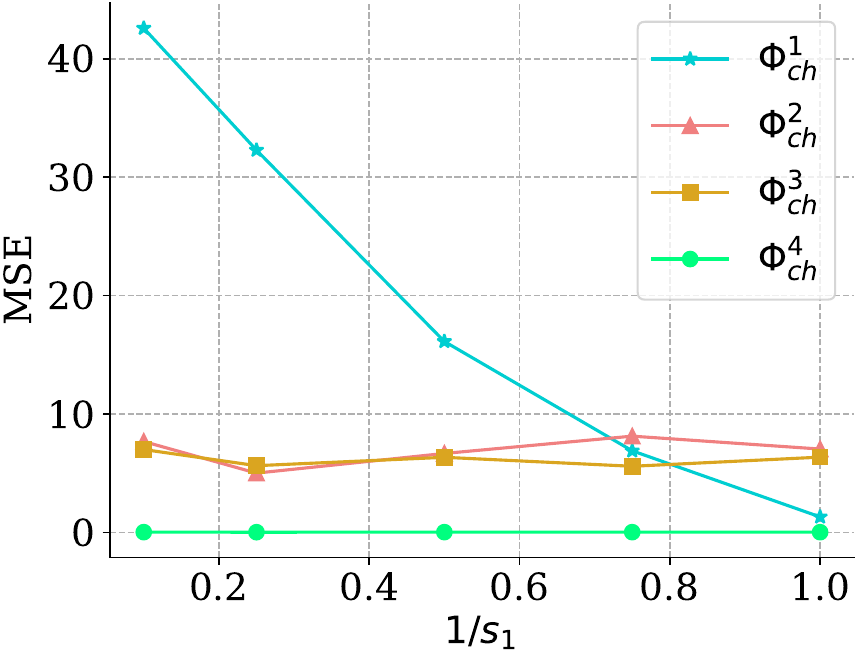}}
  \centerline{(a)}\medskip
\end{minipage}
\hfill
\begin{minipage}[b]{0.49\linewidth}
  \centering
  \centerline{\includegraphics[width=\linewidth]{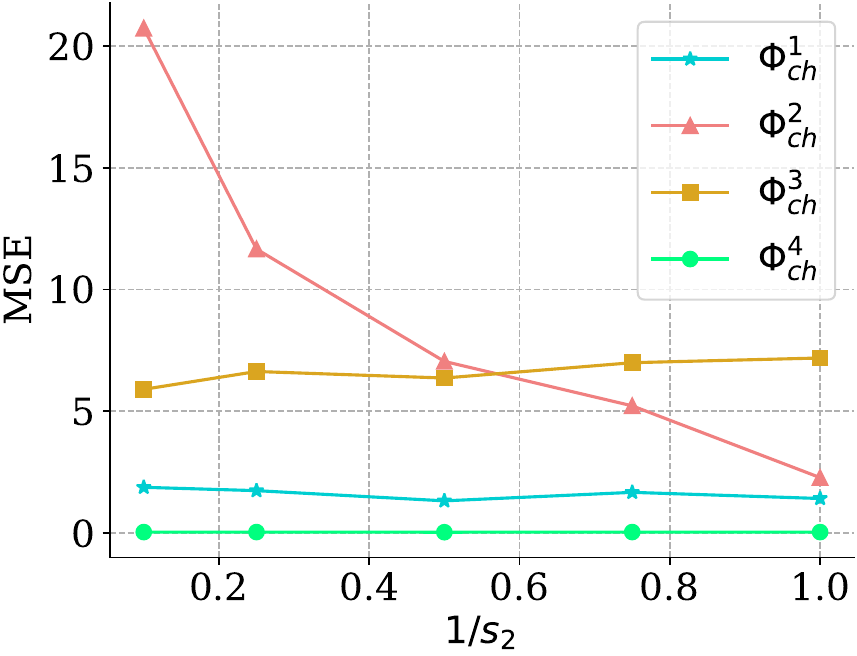}}
  \centerline{(b)}\medskip
\end{minipage}

\begin{minipage}[b]{0.49\linewidth}
  \centering
  \centerline{\includegraphics[width=\linewidth]{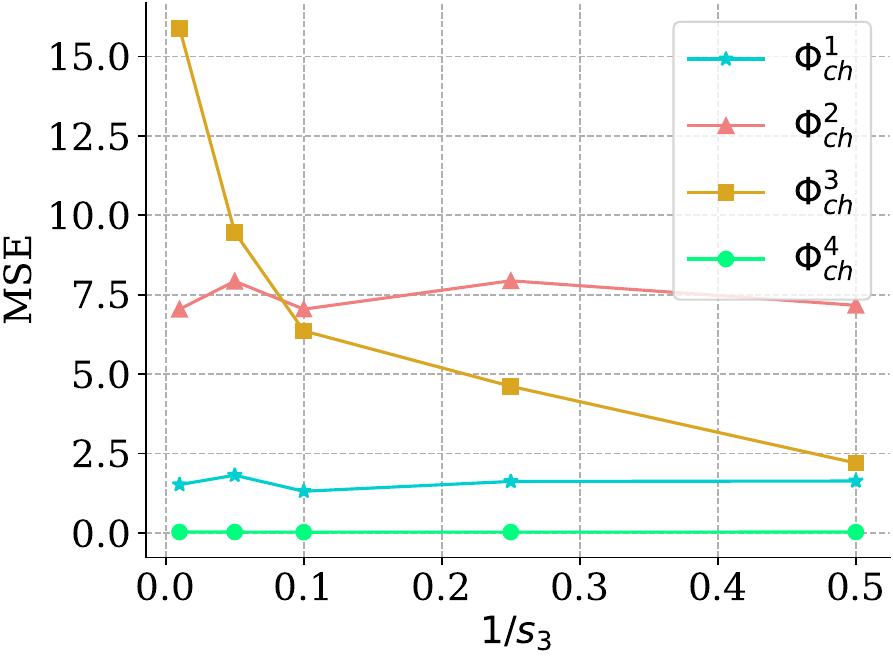}}
  \centerline{(c)}\medskip
\end{minipage}
\hfill
\begin{minipage}[b]{0.49\linewidth}
  \centering
  \centerline{\includegraphics[width=0.95\linewidth]{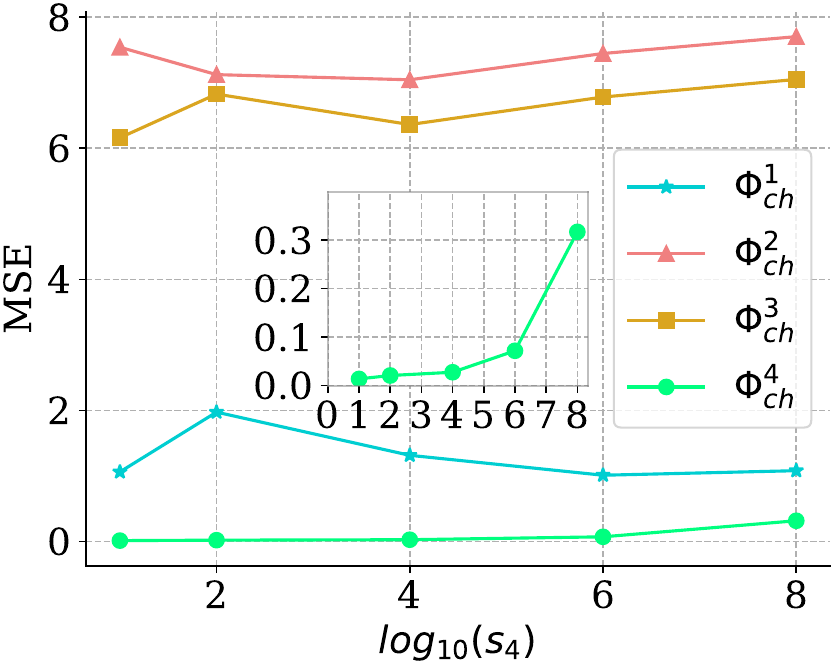}}
  \centerline{(d)}\medskip
\end{minipage}
\caption{$MSE(\mathbf{\Phi}_{\text{ch-org}}^i, \mathbf{\Phi}_{\text{ch}}^i)$ with different $s_i$. Note that $s_i$, $i\in [1,2,3]$ are plotted in $1/s_i$, and $s_4$ are plotted in $log_{10}(s_4)$ for better observation. (a) $s_1$, (b) $s_2$, (c) $s_3$, (d) $s_4$. }
\label{fig:s scale parameters}
\end{figure}

To determine the optimal scaling table $\{{s}_i\}$, we analyze the impacts of changing scaling factors in the entropy coding. We first let $s_{i-1} \leq s_i$ to ensure that larger scaling factors are assigned to groups of lesser importance. Then, we analyze the impacts of changing scaling factors in the entropy coding. We established baseline priors $\mathbf{\Phi}_{\text{ch-org}}^i$ generated by the prior extraction network $g_{\text{ch}}^i$ without any scaling. The scale table $\{{s}_i\}$ is initialized as $[1, 1, 2, 10, 1 \times 10^{4}]$, complying with the principles $s_{i-1} \leq s_i$. Then, the scaling factor ${s}_i$ was individually adjusted while other ${s}_j$, $j \neq i$ were fixed. To evaluate the impacts, the Mean Square Error (MSE) between the scaled priors $\mathbf{\Phi}_{\text{ch}}^i$ from $\mathbf{y}_N$ and the original prior $\mathbf{\Phi}_{\text{ch-org}}^i$ from $\mathbf{y}$ was calculated, which is $MSE(\mathbf{\Phi}_{\text{ch-org}}^i, \mathbf{\Phi}_{\text{ch}}^i)=||\mathbf{\Phi}_{\text{ch-org}}^i-\mathbf{\Phi}_{\text{ch}}^i||^2$.

Fig.~\ref{fig:s scale parameters} shows the $MSE(\mathbf{\Phi}_{\text{ch-org}}^i, \mathbf{\Phi}_{\text{ch}}^i)$ when $\mathbf{y}$ is scaled with different factors $s_i$. We can have the following four observations: 1) As $\{{s}_i\}$ is initialized $[1, 1, 2, 10, 1 \times 10^{4}]$, there is a mismatch between $\mathbf{\Phi}_{\text{ch-org}}^i$ and $\mathbf{\Phi}_{\text{ch}}^i$, i.e., $MSE(\mathbf{\Phi}_{\text{ch-org}}^i, \mathbf{\Phi}_{\text{ch}}^i)$ are non-zero. 
2) When changing the scaling factor $s_i$, the MSE of priors $\mathbf{\Phi}_{\text{ch}}^i$, $MSE(\mathbf{\Phi}_{\text{ch-org}}^i, \mathbf{\Phi}_{\text{ch}}^i)$, decreases significantly, which indicates that $s_i$ has high impacts on the priors $\mathbf{\Phi}_{\text{ch}}^i$. 3) The rest of priors $\mathbf{\Phi}_{\text{ch}}^j$, $j\neq i$ maintain almost the same, which indicates that $s_i$ has little impact on the priors $\mathbf{\Phi}_{\text{ch}}^j$, $j\neq i$. 4) In Fig.~\ref{fig:s scale parameters}(d), it is observed that the $ \mathbf{\Phi}_{\text{ch}}^4$ increases slightly as $log_{10}(s_4)$ increases. In addition, other $ \mathbf{\Phi}_{\text{ch}}^i$, $i\neq4$ vary randomly. 
Based on these findings, the scaling factor $s_i$ is able to control the accuracy of $\mathbf{\Phi}_{\text{ch}}^i$ and $s_i$ works independently. Thus, we optimize the factors one-by-one from its initial values $[1, 1, 2, 10, 1 \times 10^{4}]$ to find their optimals. 

\begin{figure}
\begin{minipage}[b]{0.49\linewidth}
  \centering
  \centerline{\includegraphics[width=\linewidth]{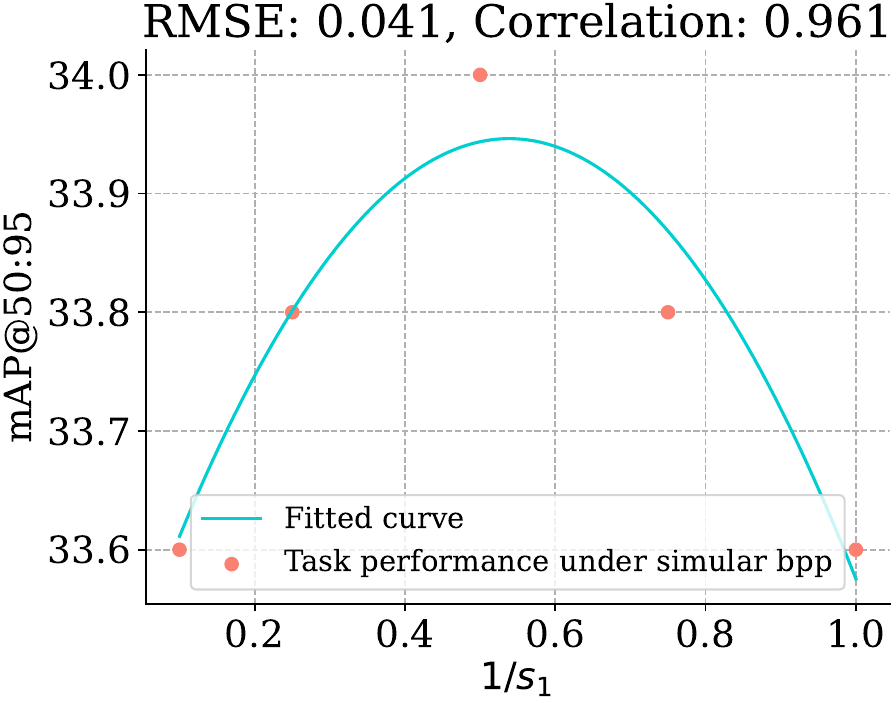}}
  \centerline{(a)}\medskip
\end{minipage}
\hfill
\begin{minipage}[b]{0.49\linewidth}
  \centering
  \centerline{\includegraphics[width=\linewidth]{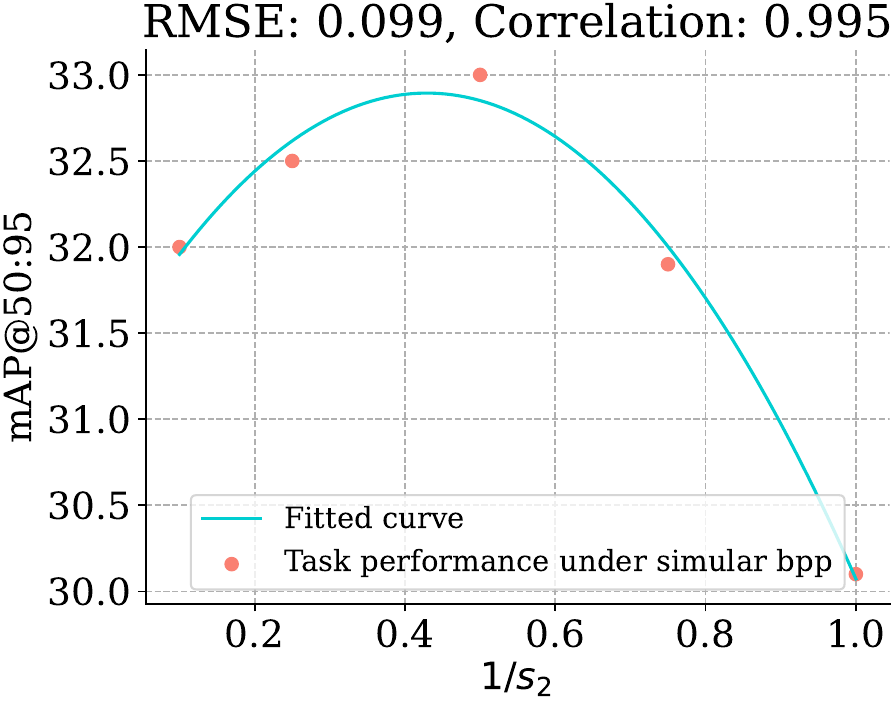}}
  \centerline{(b)}\medskip
\end{minipage}

\begin{minipage}[b]{0.49\linewidth}
  \centering
  \centerline{\includegraphics[width=0.98\linewidth]{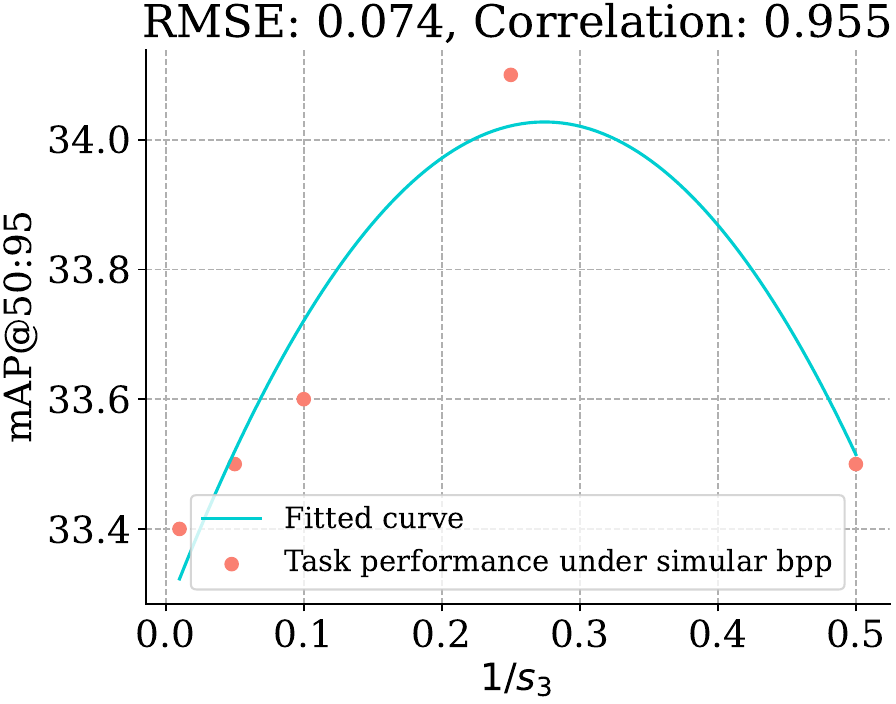}}
  \centerline{(c)}\medskip
\end{minipage}
\hfill
\begin{minipage}[b]{0.49\linewidth}
  \centering
  \centerline{\includegraphics[width=\linewidth]{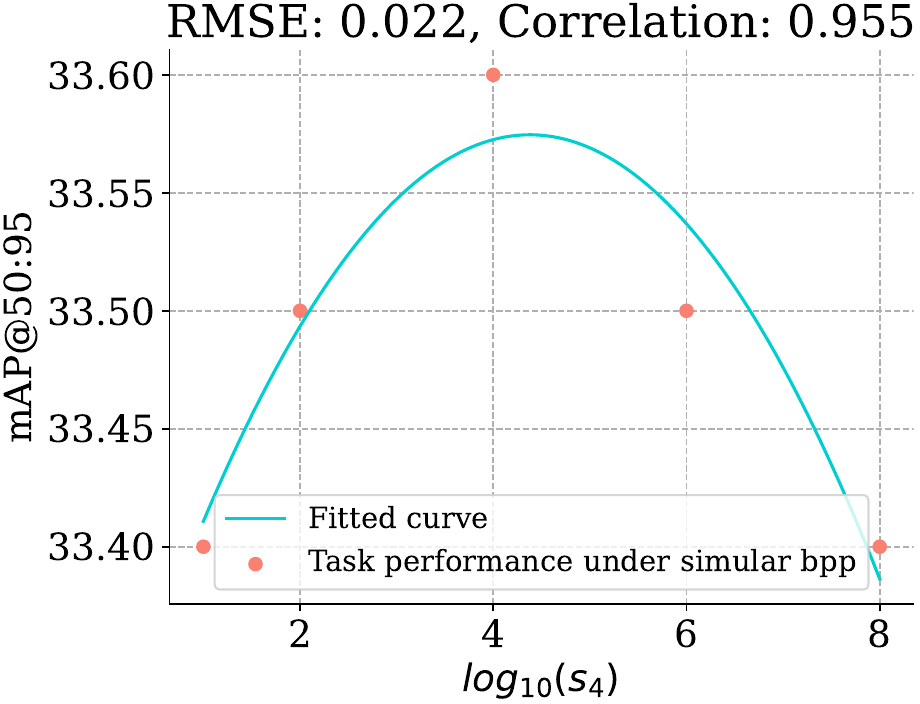}}
  \centerline{(d)}\medskip
\end{minipage}
\caption{Object task precisions using different $s_i$, where quadratic functions are used to fit results. Note that $s_i$, $i\in [1,2,3]$ are plotted in $1/s_i$, and $s_4$ are plotted in $log_{10}(s_4)$ for better observation. (a) $s_1$, (b) $s_2$, (c) $s_3$, (d) $s_4$.}
\label{fig:scale table}
\end{figure}

Starting with the initialized table, we fine-tuned the scale factors and the base ICM codec to achieve high machine vision task accuracy at a given bit rate. For the primary object detection task, Faster R-CNN was used as a task model, and mAP@50:95 was used to evaluate the detection accuracy. We performed coding experiments on the COCO2017 validation dataset. Fig.~\ref{fig:scale table} shows the changes in mAP@50:95 while using different $s_i$, where the red dots are real data and the blue lines are fitting curves with quadratic functions. We can observe that the Root MSE (RMSE) of the fitting error is from 0.041 to 0.099, which are small. The Correlation Coefficients (CC) of the fitting curves are 0.955 to 0.995, which is high. The high CC and small RMSE show that the quadratic function is accurate. Therefore, based on these fitted curves, the optimal $s_i$ is obtained to achieve the highest mAP@50:95, which is $[1, 1.85, 2.27, 3.71, 1 \times 10^{4.38}]$. Note that these optimal values are obtained from COCO2017 validation dataset and object detection, and are kept the same under different settings. Their generalization capability for other tasks, such as segmentation and dataset, will be investigated in the experimental section.


\begin{figure}
\centering
\centerline{\includegraphics[width=0.65\linewidth]{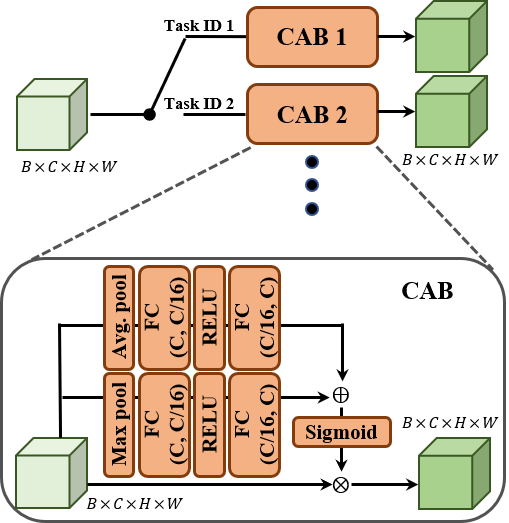}}
\caption{Structure of the TSCA module.}
\label{fig:tsca}
\end{figure}
\subsection{Proposed TSCA Module}
\label{tsca}
To adapt to multiple machine tasks, we develop TSCA module in the proposed CI-ICM. Fig.~\ref{fig:tsca} shows the structure of the TSCA module, which includes a bank of parallel Channel Attention Blocks (CABs) and each CAB specialized for a particular downstream task, e.g., object detection and instance segmentation. Once the downstream task is identified, the TSCA is set to the corresponding task ID, {and only one CAB is activated. For a new task, a new CAB is trained and integrated into the TSCA to generate the optimal bitstream.} Within each CAB, features first undergo global max pooling and global average pooling along the spatial dimension, adopting the channel adaptation mechanism. This designation extracts the mean and maximum values per channel, generating two feature vectors of length 128, which match the channel count of our network. These vectors are then transformed through a shared network of two fully connected layers with an intermediate ReLU activation function. The outputs of these shared networks are summed and passed through a sigmoid activation function to produce channel-wise enhancement factors, denoted as $\gamma_{i}^{\text{TSCA}_k}$ for $i$-th channel in $k$-th TSCA module. These factors are used to weight the original features via channel-wise self-attention, prioritizing the important channels for the specific task.

The CAB uses self-attention to enhance task-specific features by integrating both mean and maximum statistics for a comprehensive channel characterization. Within each TSCA module, a channel order loss is applied to the enhancement factors, guiding the codec to learn an ordered and effective feature representation during training. Consequently, the TSCA module adaptively enhances relevant features for diverse tasks, while integrated loss ensures that the network extracts and organizes those features effectively. 

\begin{figure}
\centering
\centerline{\includegraphics[width=0.8\linewidth]{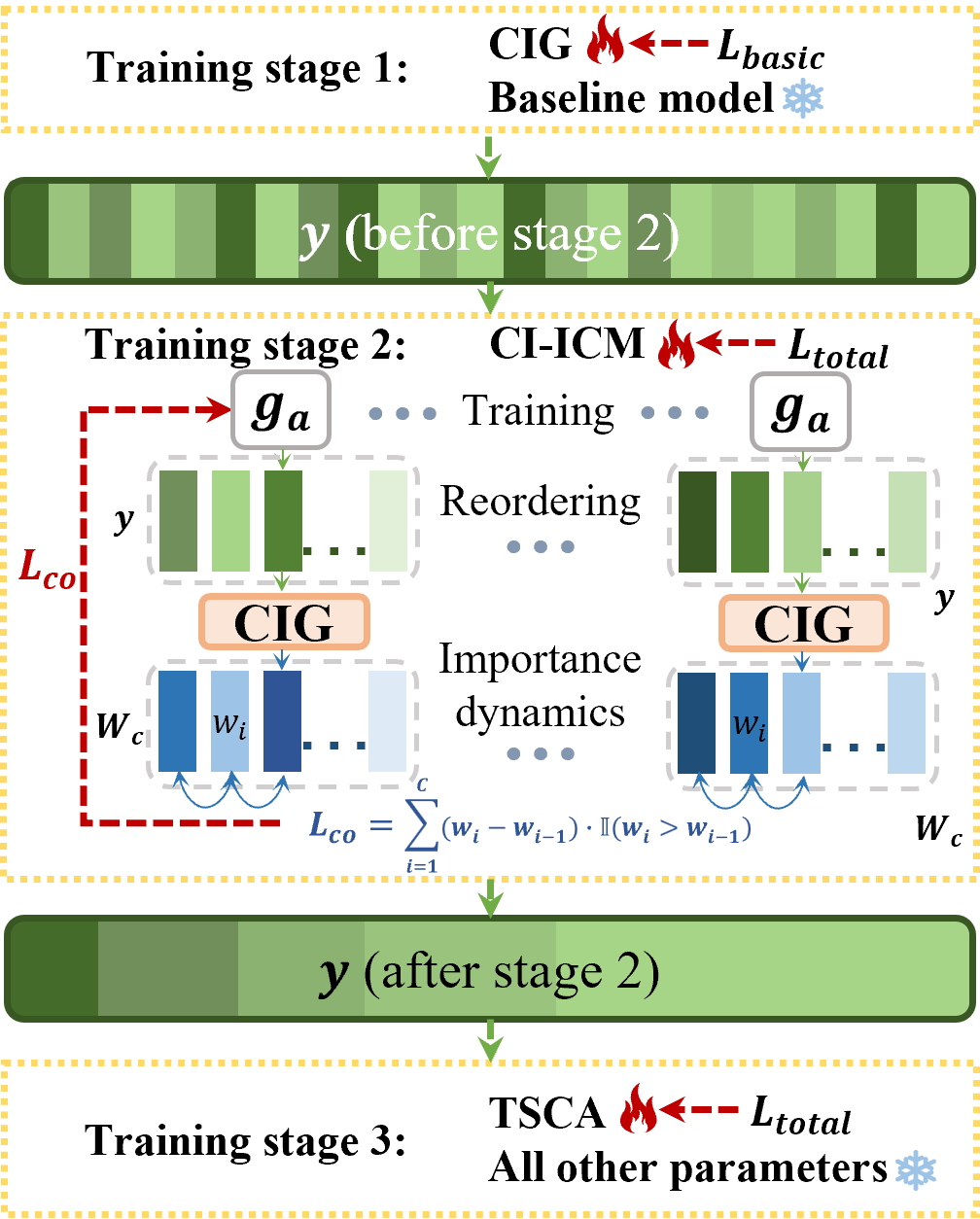}}
\caption{Three stages of training for the proposed CI-ICM, where dark green $\mathbf{y}$ blocks are features with higher importance for machine vision, and dark blue $\mathbf{W}_{c}$ blocks are the higher importance weights.}
\label{fig:ff}
\end{figure}
\subsection{Loss Functions and Training Procedures}
We first present the channel order loss for feature extraction, the loss functions for model training, and the training procedure of the proposed CI-ICM.

\subsubsection{Channel Order Loss}
\label{COL}
To extract representative features for machine vision, we designed a channel order loss $L_{CO}$, which guides the encoder $g_a$ to organize the feature channels in $\mathbf{y}$ according to a descending order of importance. Inspired by the monotonicity loss in~\cite{monotonicity_loss}, $L_{CO}$ is formulated to penalize any pair of adjacent channels where the importance weight of a subsequent channel is greater than its predecessor. $L_{CO}$ consists of two key parts. Let $L_{co}^{\text{CIG}}$ denote channel order loss computed in CIG, from which $\mathbf{W}_{c}$ is generated, and $w_{i}^{\text{CIG}}$ denote the importance weight of $i$-th channel in $\mathbf{W}_{c}$. If $w_{i}^{\text{CIG}}-w_{i-1}^{\text{CIG}}>0$, this constitutes a misarrangement, and the difference is accumulated. Otherwise, no loss is incurred. The $L_{co}^{\text{CIG}}$ based on $\mathbf{W}_{c}$ is calculated as
\begin{equation}
L_{co}^{\text{CIG}} = \sum_{i=1}^{C} (w_{i}^{\text{CIG}} - w_{i-1}^{\text{CIG}}) \cdot \mathbb{I}(w_{i}^{\text{CIG}} > w_{i-1}^{\text{CIG}}),
\label{channel order loss cig}
\end{equation}
where $C$ denotes the number of channels in the latent representation $\mathbf{y}$, and $\mathbb{I}(\cdot)$ denotes the indicator function which yields 1 if the condition is true. Otherwise 0.

Inspired by \cite{Rate-DistortionTheory2025}, the feature extraction process in $g_a$ progressively discards redundant information to refine machine vision-oriented features. We propose using a channel order loss to guide the extraction of ordered features. Given that TSCA generates channel enhancement factors to adapt features for specific tasks, we compute the channel order loss of TSCA as
\begin{equation}
L_{co}^{\text{TSCA}_k} = \sum_{i=1}^{C_k} (\gamma_{i}^{\text{TSCA}_k} - \gamma_{i-1}^{\text{TSCA}_k}) \cdot \mathbb{I}(\gamma_{i}^{\text{TSCA}_k} > \gamma_{i-1}^{\text{TSCA}_k}),
\label{channel order loss tsca}
\end{equation}
where $L_{co}^{\text{TSCA}_k}$ is the loss calculated in the $k$-th TSCA module with $\gamma_{i}^{\text{TSCA}_k}$ denoting the enhancement factor of $i$-th channel, and $C_k$ denotes the number of channels in $\text{TSCA}_k$. The total channel order loss
$L_{CO}$ consists of $L_{co}^{\text{CIG}}$ and $L_{co}^{\text{TSCA}_k}$.

\subsubsection{Loss Functions for Compression Network}
\label{training loss functions}
The training loss of our CI-ICM involves two loss functions, which are the basic loss function and the total loss function enhanced with channel order loss. The basic loss function includes feature distortion and bitrate, where the loss of the MSE is calculated in the feature space of the task network and serves as the distortion loss. The basic loss function for feature coding is formulated as
\begin{equation}
L_{basic} = \underbrace{{D}(\mathbf{F}, \mathbf{F}')}_{\text{Feature distortion loss}} + \underbrace{\lambda (R_{\hat{\mathbf{y}}}+R_{\hat{\mathbf{z}}})}_{\text{Rate loss}}.
\label{loss1}
\end{equation}
where $\mathbf{F}$ and $\mathbf{F}'$ represent the intermediate features obtained from the machine vision task network using the original and compressed images, respectively. $D()$ measures the feature difference, which uses MSE. $\lambda$ is the Lagrange multiplier controlling the trade-off between bit rate and distortion. 
$R_{\hat{\mathbf{y}}}$ and $R_{\hat{\mathbf{z}}}$ denote the bitrate of the decoded feature $\hat{\mathbf{y}}$ and the decoded side information $\hat{\mathbf{z}}$.
The rate-loss terms are estimated by the entropy model based on CI-CTX, which is 
\begin{equation}
\begin{cases}
    & R_{\hat{\mathbf{y}}} = \sum_{i=1}^{n} R_{\hat{\mathbf{y}}^i} \\
    & R_{\hat{\mathbf{y}}^0} = \mathbb{E}[ -log_{2}( p_{\hat{\mathbf{y}}^0 | \mathbf{\Phi}_{\text{z}}^0}(\hat{\mathbf{y}}^0 | \mathbf{\Phi}_{\text{z}}^0 ))] \\
    & R_{\hat{\mathbf{y}}^i} = \mathbb{E}[ -log_{2}( p_{\hat{\mathbf{y}}^i | \mathbf{\Phi}_{\text{ch}}^i} (\hat{\mathbf{y}}^i | \mathbf{\Phi}_{\text{ch}}^i))]\\
    & R_{\hat{\mathbf{z}}} = \mathbb{E}[-log_{2}(p_{\hat{\mathbf{z}}}(\hat{\mathbf{z}}))]
\end{cases},
\end{equation}
where $R_{\hat{\mathbf{y}}^i}$ denote the bitrate of the $i$-th feature subset, $n$ the the total number of subsets, $\mathbf{\Phi}_{\text{z}}^0$ and $\mathbf{\Phi}_{\text{ch}}^i$  denotes the hyper-prior parameters of $i$-th feature channels.
To enforce ordered feature extraction and proper feature separation, channel order loss is added to formulate the total loss for the proposed CI-ICM framework, which is
\begin{equation}
L_{total} = L_{basic} + \underbrace{(\sum_{k=1}^{N} \varphi_{k} \cdot L_{co}^{\text{TSCA}_k}) + \varphi_{CIG}L_{co}^{\text{CIG}}}_{\text{Channel Order Loss }L_{CO}},
\label{loss2}
\end{equation}
where $L_{basic}$ is the basic coding loss illustrated in Eq. \ref{loss1}. In $L_{co}$, $L_{co}^{\text{CIG}}$ and $L_{co}^{\text{TSCA}_k}$ are the proposed channel order losses calculated in the $k$-th TSCA module and the CIG module, $N$ is the number of TSCAs in the encoder and $N$ is set as 3 in our CI-ICM, $\varphi_{k}$ and $\varphi_{CIG}$ are the weighting factors controlling the impact of channel order losses.

\begin{figure}
\begin{minipage}[b]{0.49\linewidth}
  \centering
  \centerline{\includegraphics[width=\linewidth]{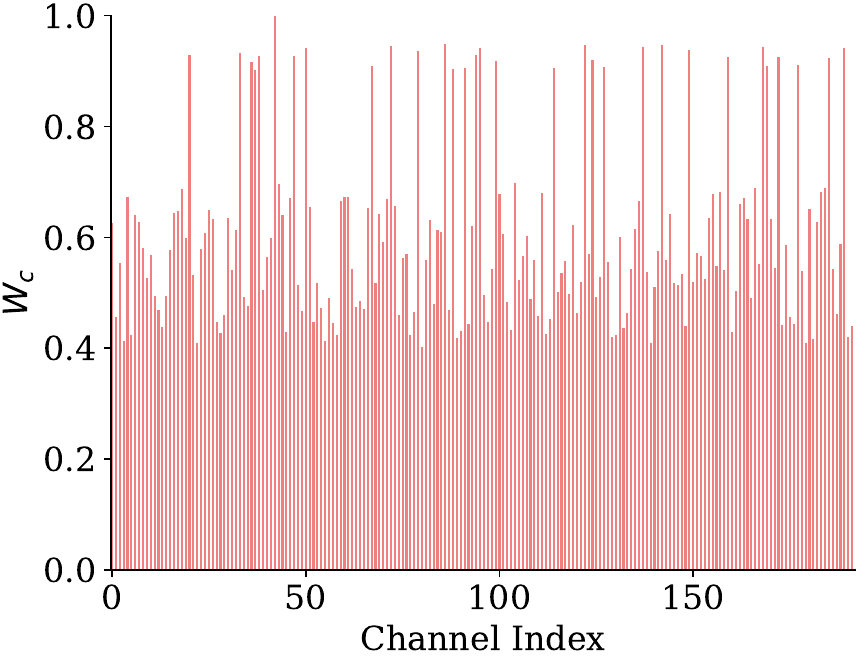}}
  \centerline{(a)}\medskip
\end{minipage}
\hfill
\begin{minipage}[b]{0.49\linewidth}
  \centering
  \centerline{\includegraphics[width=\linewidth]{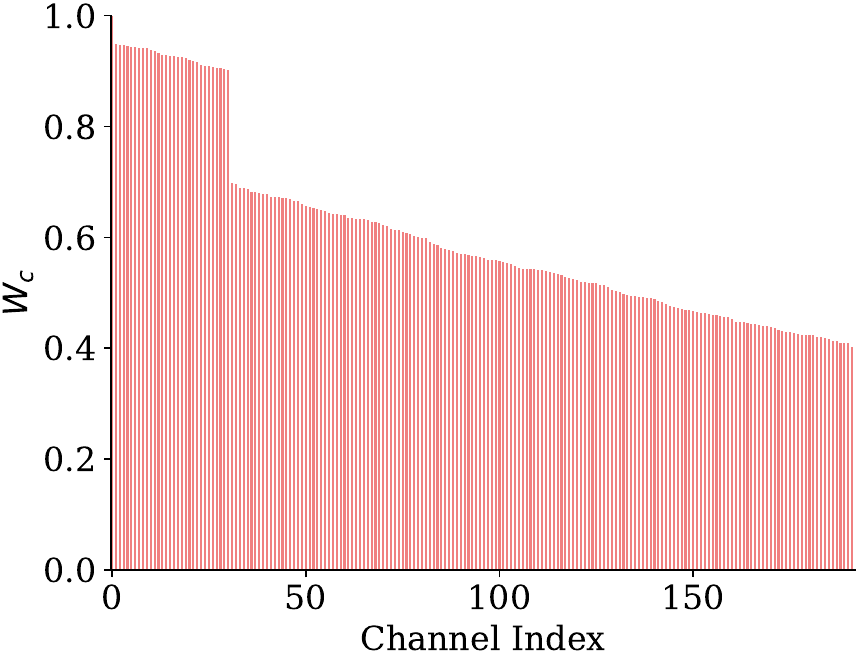}}
  \centerline{(b)}\medskip
\end{minipage}
\caption{Trained $\mathbf{W}_{c}$ of an image in COCO2017 before and after training using channel order loss. (a) Before training, (b) After training.}
\label{fig:coloss}
\end{figure}

\subsubsection{Training Procedure}
\label{lf}
The training procedure for the proposed CI-ICM consists of three stages with three key objectives: feature importance evaluation, ordered representation generation with bitrate allocation, and multi-task adaptation, as shown in Fig.~\ref{fig:ff}. Notably, we freeze the machine vision task model at all stages. In stage 1, we only introduced the CIG module to the baseline TransTIC compression model \cite{Chen2023TransTICTT} for the evaluation of channel importance, and initialized the baseline TransTIC model with checkpoints of \cite{Chen2023TransTICTT}. The training process begins with independent optimization of the CIG module while the baseline TransTIC model is frozen. $L_{basic}$ is used as the loss function to train CIG and analyze feature channel importance in stage 1. This stage is critical for establishing a stable and efficient feature importance estimator to provide a robust starting point for the joint training. 

Stage 2 involves an end-to-end joint training of the entire CI-ICM framework, which simultaneously optimizes the encoder $g_a$ to extract features $\mathbf{y}$ in descending order of importance and employs FCGS and CI-CTX for bitrate allocation. {The scaling factors used in FCGS are the optimal scaling table determined in Section \ref{sm_2}, which are $[1, 1.85, 2.27, 3.71, 1 \times 10^{4.38}]$.} $L_{total}$ is used as the loss function for CI-ICM modules in stage 2. As shown in Fig.~\ref{fig:ff} and Fig.~\ref{fig:pipeline}(b) ``Reordering", initially randomly ordered features are automatically in descending importance after training. To validate the effectiveness of feature reordering, Fig.~\ref{fig:coloss} illustrates the channel importance weights $\mathbf{W}_{c}$ from CIG before and after stage 2 training.
It is observed that features from $g_a$ are randomly ordered in Fig.~\ref{fig:coloss}(a), which are placed in descending order in Fig.~\ref{fig:coloss}(b) after training. The proposed channel order loss is effective. Moreover, other coding modules in CI-ICM are also trained with $L_{total}$. Stage 2 is the main training phase for CI-ICM, which aims to maximize ICM coding performance.

Stage 3 is to adapt the TSCA modules for multi-task scenarios. During this stage, only the lightweight TSCA modules are trained and updated by incorporating new CABs, while all other components in the CI-ICM are kept frozen. The loss function $L_{total}$ is employed, which is identical to that of stage 2. This approach mitigates the need for costly full-scale retraining for each new task, enabling rapid and cost-effective adaptation, making the proposed CI-ICM highly suitable for multitask deployments. Through this three-stage training, the CI-ICM achieves both high coding efficiency and robust multi-task adaptation to real-world machine vision applications.

\section{Experimental Results and Analysis}
\label{exp results and analysis}
This section presents experimental results and analysis of the proposed CI-ICM, including experimental settings, coding performance evaluation on two machine tasks, ablation studies on key modules, {generalization analysis}, and computational complexity analysis.

\begin{figure*}[!ht]
\begin{minipage}[b]{0.28\linewidth}
  \centering
  \centerline{\includegraphics[width=\linewidth]{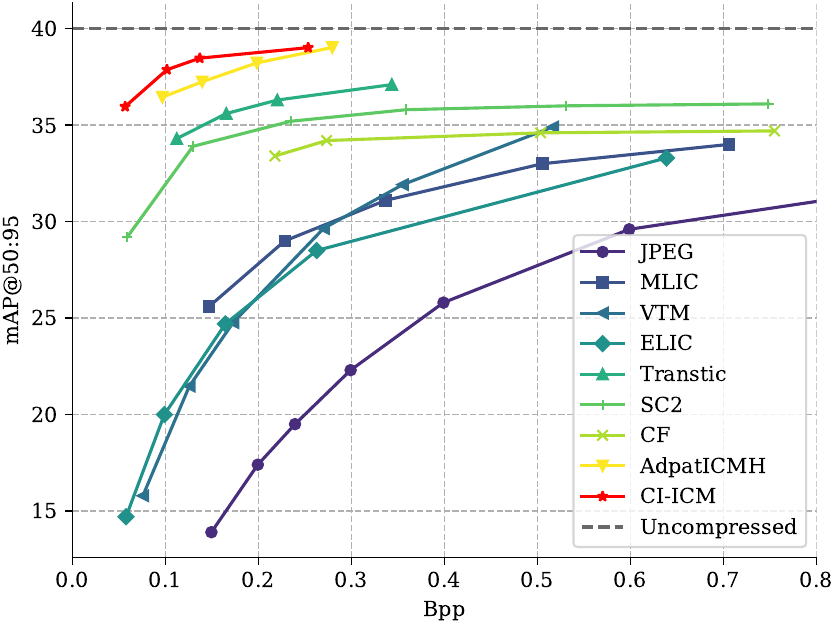}}
  \centerline{(a)}\medskip
\end{minipage}
\hfill
\begin{minipage}[b]{0.28\linewidth}
  \centering
  \centerline{\includegraphics[width=\linewidth]{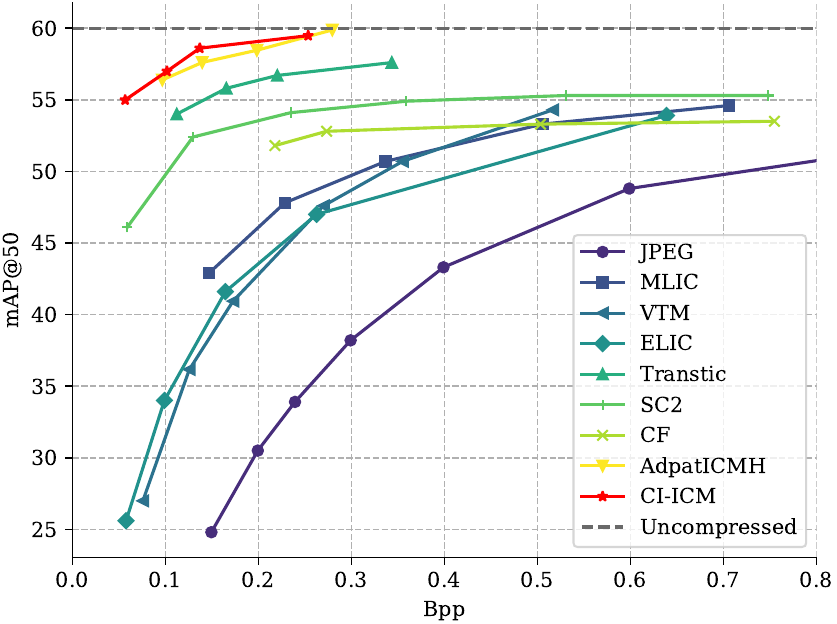}}
  \centerline{(b)}\medskip
\end{minipage}
\hfill
\begin{minipage}[b]{0.28\linewidth}
  \centering
  \centerline{\includegraphics[width=\linewidth]{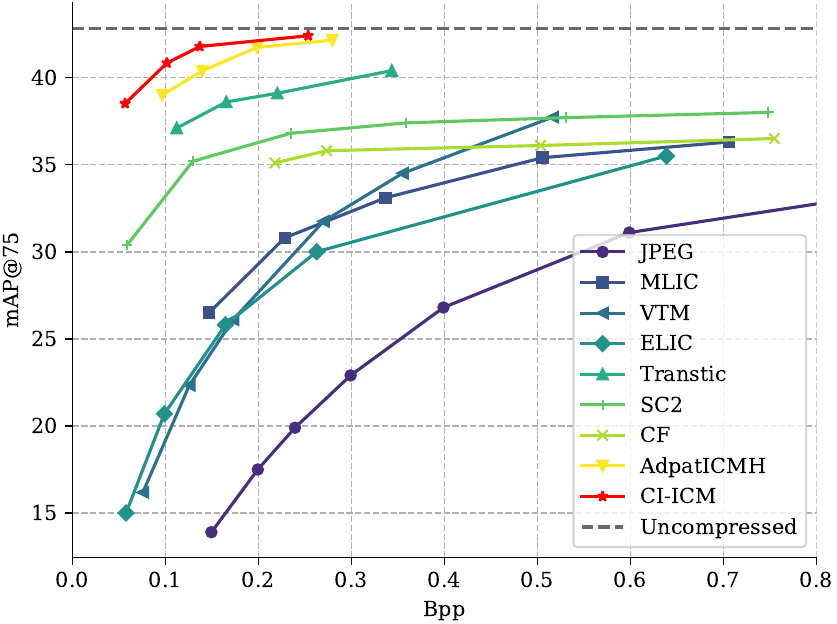}}
  \centerline{(c)}\medskip
\end{minipage}
\caption{Rate-accuracy curves of the proposed CI-ICM and benchmark schemes on object detection task, (a) mAP@50:95, (b) mAP@50, (c) mAP@75.}
\label{fig:objdet}
\end{figure*}

\subsection{Experimental Setup}
\label{expsetup}

To comprehensively evaluate the performance of the CI-ICM framework, the experimental protocols are designed and presented in detail. We establish a practical multitask scenario where object detection serves as the primary task, executed by Faster R-CNN \cite{fasterrcnn}, and instance segmentation acts as the secondary task performed by Mask R-CNN \cite{He2017MaskR}. Both task networks are common test conditions in VCM.

\subsubsection{Benchmark Coding Schemes}
We selected typical human vision-oriented compression methods and state-of-the-art ICM methods as benchmark schemes for comparison. For the primary object detection task, the human vision-oriented methods include: the JPEG2000~\cite{skodras2001jpeg} denoted as ``JPEG", {the VTM version 23.13~\cite{vvc} denoted as ``VTM",} the state-of-the-art pretrained learned image compression methods~\cite{He2022ELICEL} labeled as ``ELIC”, and~\cite{Jiang2022MLICME} labeled as ``MLIC", which are officially provided by authors. The ICM baseline methods include: the supervised compression method~\cite{matsubara2023sc} labeled as ``SC2”, the end-to-end feature compression method~\cite{end-to-end} labeled as ``CF”, the prompt tuning based ICM~\cite{Chen2023TransTICTT} labeled as ``TransTIC”, and the spatial-frequency adaptation based ICM~\cite{Li2024ImageCF} labeled as ``AdaptICMH”, which is the state-of-the-art ICM. 
For the secondary instance segmentation task, benchmark schemes include human-centric image compression methods: JPEG~\cite{skodras2001jpeg}, {VTM~\cite{vvc},} ELIC~\cite{He2022ELICEL}, MLIC~\cite{Jiang2022MLICME}; ICM methods: TransTIC~\cite{Chen2023TransTICTT} and AdaptICMH~\cite{Li2024ImageCF}. {SC2 and CF were excluded from the instance segmentation evaluation because they were only optimized for object detection using Faster R-CNN ResNet-50 as backbone.}

\subsubsection{Training Datasets and Configurations}
The CI-ICM was trained on a server computer with NVIDIA GeForce RTX 3090 GPUs and PyTorch platform. COCO2017 was used as the training dataset. We started from the pre-trained checkpoints \cite{Chen2023TransTICTT}, and trained models for four distinct compression levels by setting $\lambda$ as $\{2,1,0.5,0.2\}$, {where one model was trained for one target bitrate}. Training procedures are presented in Section~\ref{lf}. In the first two stages, the model is trained for 30 epochs each, using a learning rate of $1 \times 10^{-4}$ and a batch size of 12. Specifically, in stage 2, coefficients $\varphi_{k} $ and $\varphi_{CIG}$ in Eq. \ref{loss2} are set as 0.1 and 0.3. In stage 3, for multi-task adaptation, we exclusively updated the TSCA modules for 10 epochs with a learning rate of $1 \times 10^{-4}$ and a batch size of 32, while all other parameters are fixed.

\subsubsection{Testing Datasets and Evaluation Configurations}
To evaluate the coding performance of CI-ICM, the images of the COCO2017 validation data set are compressed at four quality levels. Two machine vision tasks, object detection and instance segmentation, were performed to evaluate the coded images. Three average precision metrics with IoU thresholds $50\%$, $75\%$ and ranging from $50\%$ to $95\%$, which are denoted as mAP@50, mAP@75, and mAP@50:95 \cite{miou}, respectively. In addition, the bit rate of the encoded images was measured with bits per pixel (bpp). The Bjøntegaard Delta Bit Rate (BDBR) was not computable due to the large bit rate gaps. Instead, we thus used Bjøntegaard Delta mean average precisions, denoted as ``BD-mAP@50", ``BD-mAP@75", and ``BD-mAP@50:95", respectively, to measure the accuracy improvement at equivalent bitrates, where the ELIC is the baseline in BD-mAP calculation.

\subsection{Coding Performance for Object Detection}
\label{objdetexp}

\begin{figure*}[htb]
\begin{minipage}[b]{.19\linewidth}
  \centering
  \centerline{\includegraphics[width=3.5cm, height=2.5cm]{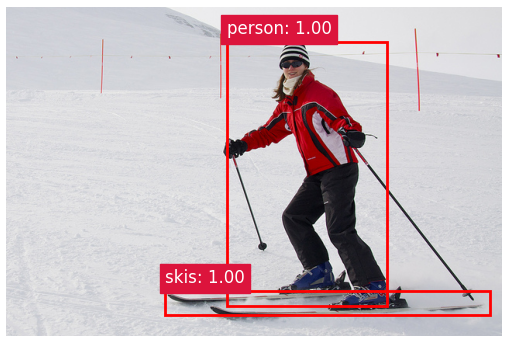}}
  \centerline{(a) Ground truth}\medskip
\end{minipage}
\hfill
\begin{minipage}[b]{.19\linewidth}
  \centering
  \centerline{\includegraphics[width=3.5cm, height=2.5cm]{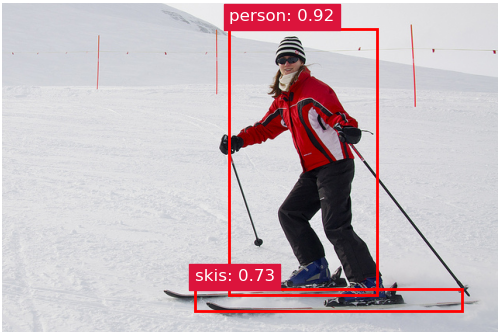}}
  \centerline{(b) ELIC (0.3613)}\medskip
\end{minipage}
\hfill
\begin{minipage}[b]{.19\linewidth}
  \centering
  \centerline{\includegraphics[width=3.5cm, height=2.5cm]{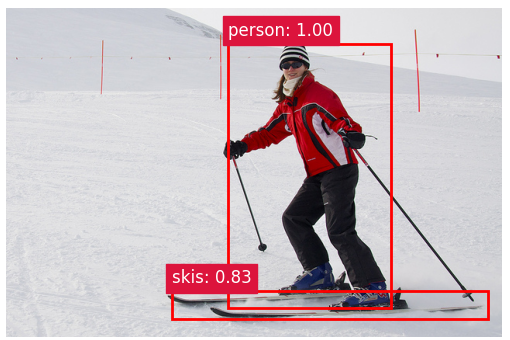}}
  \centerline{(c) TransTIC (0.2425)}\medskip
\end{minipage}
\hfill
\begin{minipage}[b]{.19\linewidth}
  \centering
  \centerline{\includegraphics[width=3.5cm, height=2.5cm]{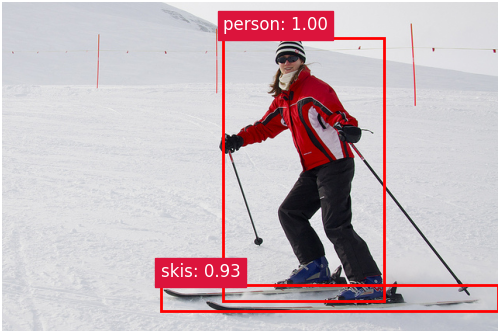}}
  \centerline{(d) AdaptICMH (0.2315)}\medskip
\end{minipage}
\hfill
\begin{minipage}[b]{.19\linewidth}
  \centering
  \centerline{\includegraphics[width=3.5cm, height=2.5cm]{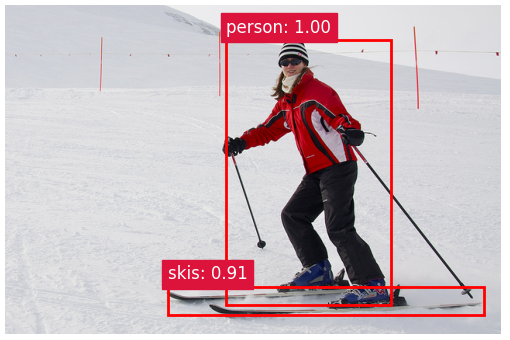}}
  \centerline{(e) CI-ICM (0.2251)}\medskip
\end{minipage}

\begin{minipage}[b]{0.19\linewidth}
  \centering
  \centerline{\includegraphics[width=3.5cm, height=3.5cm]{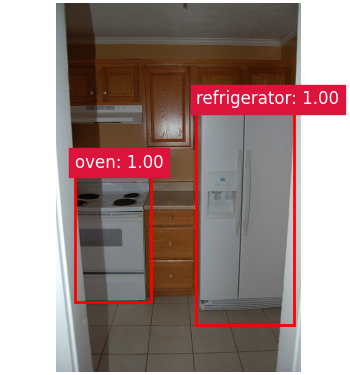}}
  \centerline{(f) Ground truth}\medskip
\end{minipage}
\hfill
\begin{minipage}[b]{0.19\linewidth}
  \centering
  \centerline{\includegraphics[width=3.5cm, height=3.5cm]{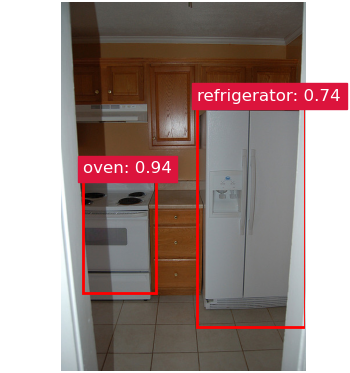}}
  \centerline{(g) ELIC (0.4012)}\medskip
\end{minipage}
\hfill
\begin{minipage}[b]{0.19\linewidth}
  \centering
  \centerline{\includegraphics[width=3.5cm, height=3.5cm]{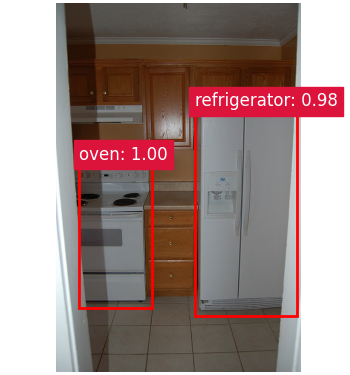}}
  \centerline{(h) TransTIC (0.2888)}\medskip
\end{minipage}
\hfill
\begin{minipage}[b]{0.19\linewidth}
  \centering
  \centerline{\includegraphics[width=3.5cm, height=3.5cm]{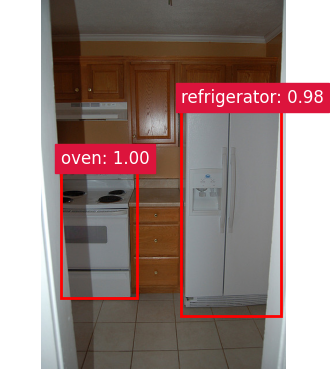}}
  \centerline{(i) AdaptICMH (0.2253)}\medskip
\end{minipage}
\hfill
\begin{minipage}[b]{0.19\linewidth}
  \centering
  \centerline{\includegraphics[width=3.5cm, height=3.5cm]{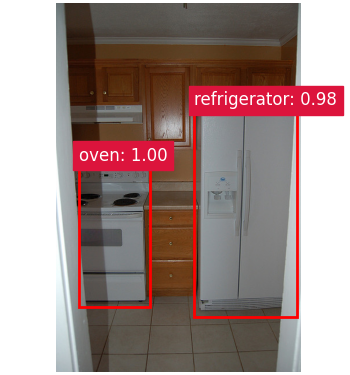}}
  \centerline{(j) CI-ICM (0.2045) }\medskip
\end{minipage}
\caption{Visualization of object detection results using different codecs, where bpp values are also presented. (a) and (f): Ground truth, (b) and (g): ELIC, (c) and (h): TransTIC, (d) and (i): AdaptICMH, (e) and (j): the proposed CI-ICM, (a)-(e): image of a person and skis, (f)-(j): image of an oven and a refrigerator.}
\label{visualization_objdet}
\end{figure*}

\renewcommand{\arraystretch}{1.5}
\begin{table}
  \setlength{\tabcolsep}{4pt}
  \caption{BD-mAPs($\%$) of the proposed CI-ICM and benchmark schemes for object detection task, where ELIC~\cite{He2022ELICEL} is the baseline.}
  \label{result table obj det}
  \begin{center}
  \begin{tabular}{c c c c}
    \hline
    Coding schemes & BD-mAP@50:95 & BD-mAP@50 & BD-mAP@75\\
    \hline
    ELIC\cite{He2022ELICEL} & 0 & 0 & 0\\
    JPEG\cite{skodras2001jpeg} & -6.798 & -9.933 & -7.729\\
    VTM\cite{vvc} & 0.171 & -0.656 & 0.511\\
    MLIC\cite{Jiang2022MLICME} & 1.128 & 1.543 & 1.418\\
    CF\cite{end-to-end} & 3.783 & 2.914 & 3.555\\
    SC2\cite{matsubara2023sc} & 8.745 & 10.348 & 8.697\\
    TransTIC\cite{Chen2023TransTICTT} & 9.864 & 12.754 & 11.542\\
    AdaptICMH\cite{Li2024ImageCF} & 13.129 & 16.845 & 15.208\\
    CI-ICM & \textbf{16.249} & \textbf{20.924} & \textbf{18.491}\\
    \hline
  \end{tabular}
  \end{center}
\end{table}
\renewcommand{\arraystretch}{1}

For object detection, {we employed Faster R-CNN with ResNet-50 as backbone.} Table \ref{result table obj det} presents the quantitative BD-mAPs of the proposed CI-ICM and the baseline methods as compared to the ELIC. The BD-mAP@50:95, BD-mAP@50, and BD-mAP@75 of JPEG are -6.798$\%$, -9.933$\%$, and -7.729$\%$, respectively, which are inferior to ELIC. {The VTM achieved BD-mAP@50:95, BD-mAP@50, and BD-mAP@75 of 0.171$\%$, -0.656$\%$, and 0.511$\%$, respectively, which are comparable to ELIC.} The BD-mAP@50:95, BD-mAP@50, and BD-mAP@75 of MLIC are enhanced by 1.128$\%$, 1.543$\%$, and 1.418$\%$, which achieved better compression performance than ELIC. Compared with ELIC, the BD-mAP@50:95 gains achieved by CF, SC2, TransTIC, and AdaptICMH are 3.783$\%$, 8.745$\%$, 9.864$\%$, and 13.129$\%$, respectively. Machine vision-oriented compression methods showed better adaptability and coding gains. The BD-mAP@50:95 BD-mAP@50 and BD-mAP@75 of the proposed CI-ICM are 16.249$\%$, 20.924$\%$, and 18.491$\%$, respectively, which are 3.120$\%$, 4.079$\%$, and 3.283$\%$ higher than those of the state-of-the-art AdaptICMH. 

Fig.~\ref{fig:objdet} illustrates the rate-accuracy curves of the proposed CI-ICM compared to the benchmark schemes. For the human-oriented image compression, traditional JPEG performs the weakest, while learned codecs ELIC and MLIC show moderate coding efficiency. ICM schemes, the CF, SC2, TransTIC, and AdaptICMH, are significantly better than the conventional human vision-oriented schemes, as they considered machine vision properties. The proposed CI-ICM has the highest accuracy at a given bit rate, which consistently exceeds all benchmark methods on object detection accuracy measured with mAP@50:95, mAP@50, and mAP@75. It owes to the consideration of channel importance. In addition,  Fig.~\ref{visualization_objdet} shows the visualization results of two compressed images, $person$ and $refrigerator$, from COCO2017 validation dataset. Compared with ELIC, TransTIC, and AdaptICMH, the proposed CI-ICM is able to maintain detection accuracies with much lower bitrates. Overall, these BD-mAPs, rate-accuracy curves, and visualized images validate that for the object detection task, the proposed CI-ICM is the best among all benchmark codecs.


\begin{figure*}[!ht]
\begin{minipage}[b]{0.28\linewidth}
  \centering
  \centerline{\includegraphics[width=\linewidth]{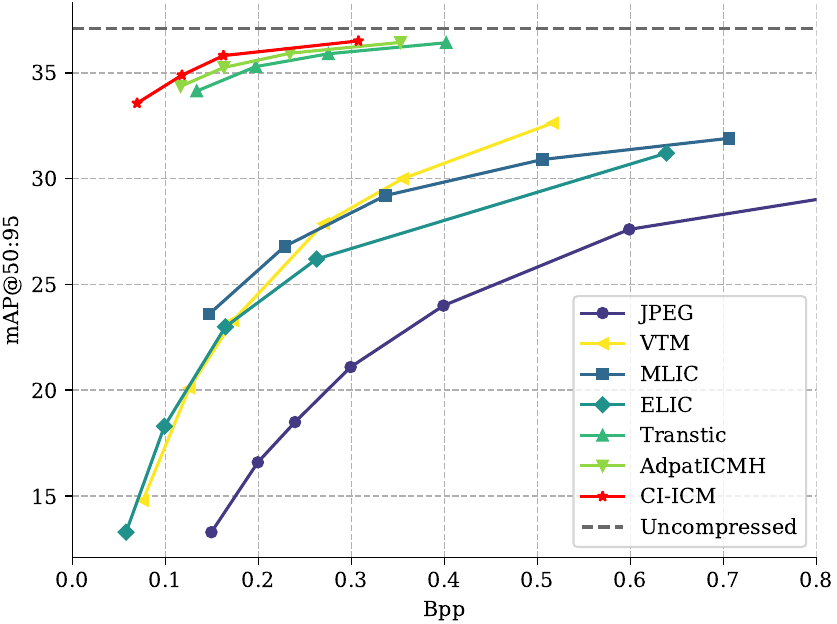}}
  \centerline{(a)}\medskip
\end{minipage}
\hfill
\begin{minipage}[b]{0.28\linewidth}
  \centering
  \centerline{\includegraphics[width=\linewidth]{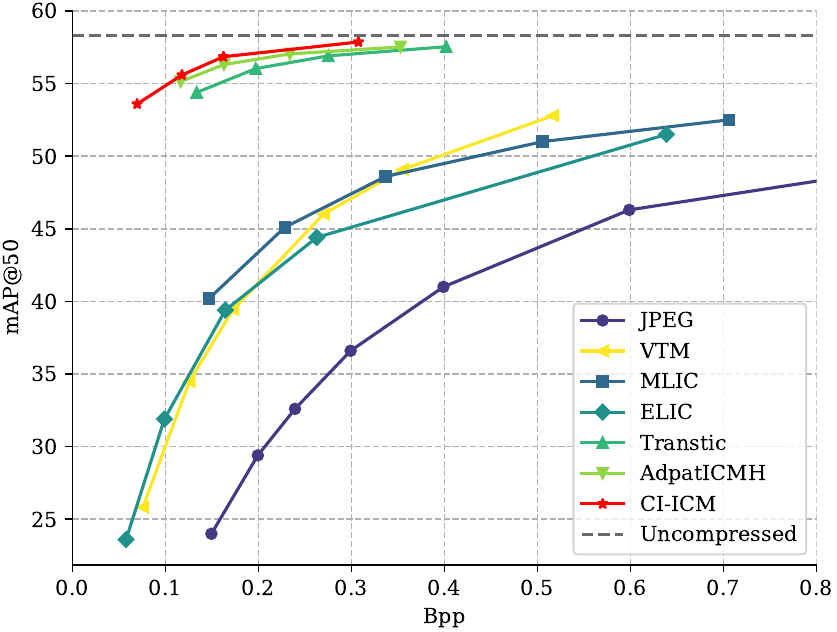}}
  \centerline{(b)}\medskip
\end{minipage}
\hfill
\begin{minipage}[b]{0.28\linewidth}
  \centering
  \centerline{\includegraphics[width=\linewidth]{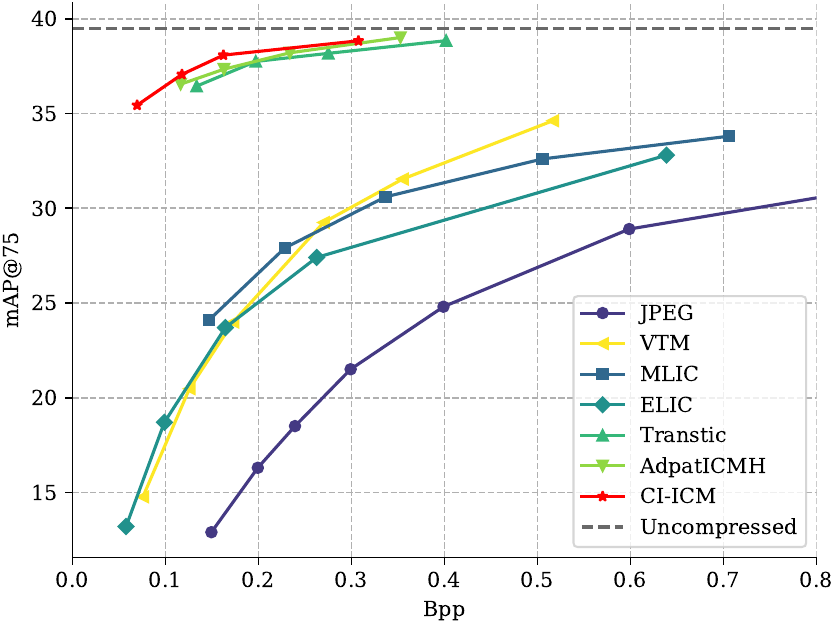}}
  \centerline{(c)}\medskip
\end{minipage}
\caption{Rate-accuracy curves of the proposed CI-ICM and baseline schemes on instance segmentation task, (a) mAP@50:95, (b) mAP@50, (c) mAP@75.}
\label{fig:insseg}
\end{figure*}

\begin{figure*}[htb]
\begin{minipage}[b]{0.19\linewidth}
  \centering
  \centerline{\includegraphics[width=3.5cm, height=2.7cm]{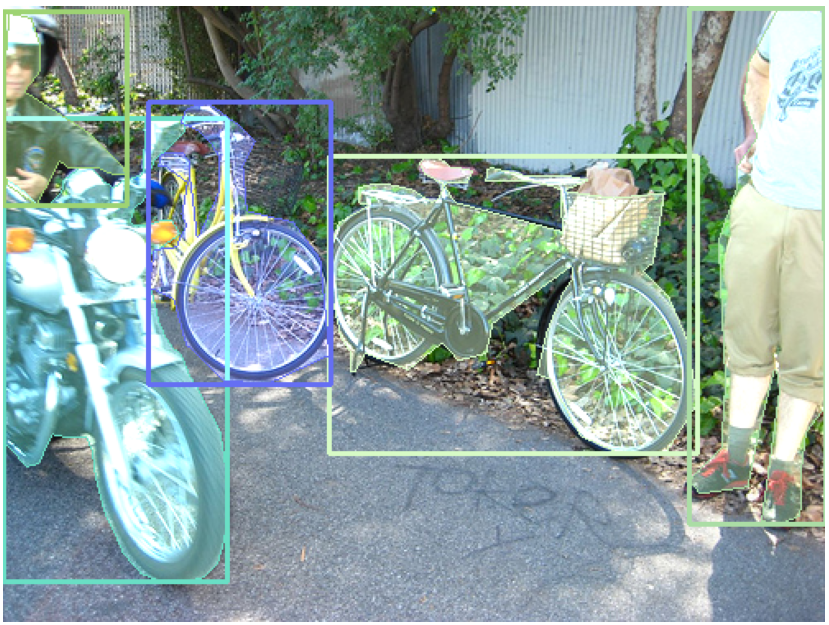}}
  \centerline{(a) Ground truth}\medskip
\end{minipage}
\hfill
\begin{minipage}[b]{0.19\linewidth}
  \centering
  \centerline{\includegraphics[width=3.5cm, height=2.7cm]{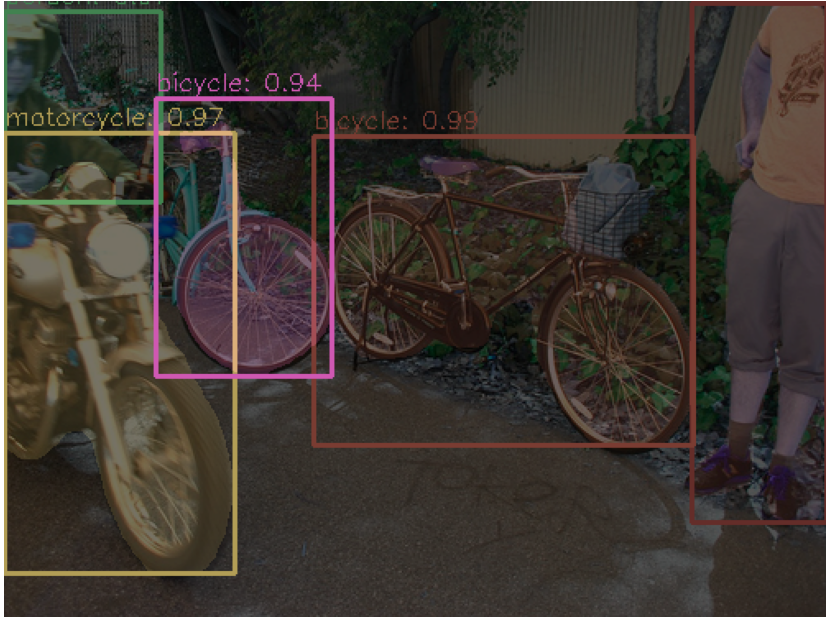}}
  \centerline{(b) ELIC (0.6924)}\medskip
\end{minipage}
\hfill
\begin{minipage}[b]{0.19\linewidth}
  \centering
  \centerline{\includegraphics[width=3.5cm, height=2.7cm]{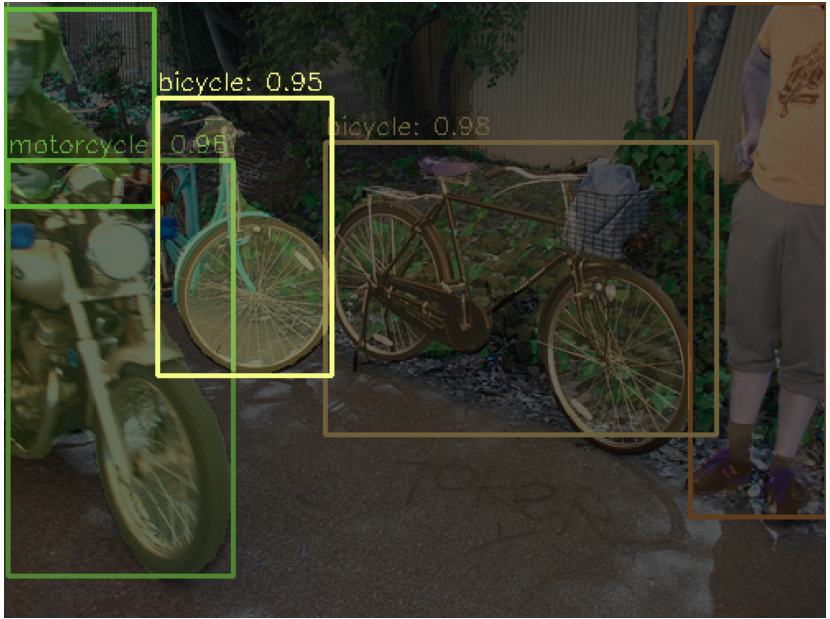}}
  \centerline{(c) TransTIC (0.5741)}\medskip
\end{minipage}
\hfill
\begin{minipage}[b]{0.19\linewidth}
  \centering
  \centerline{\includegraphics[width=3.5cm, height=2.7cm]{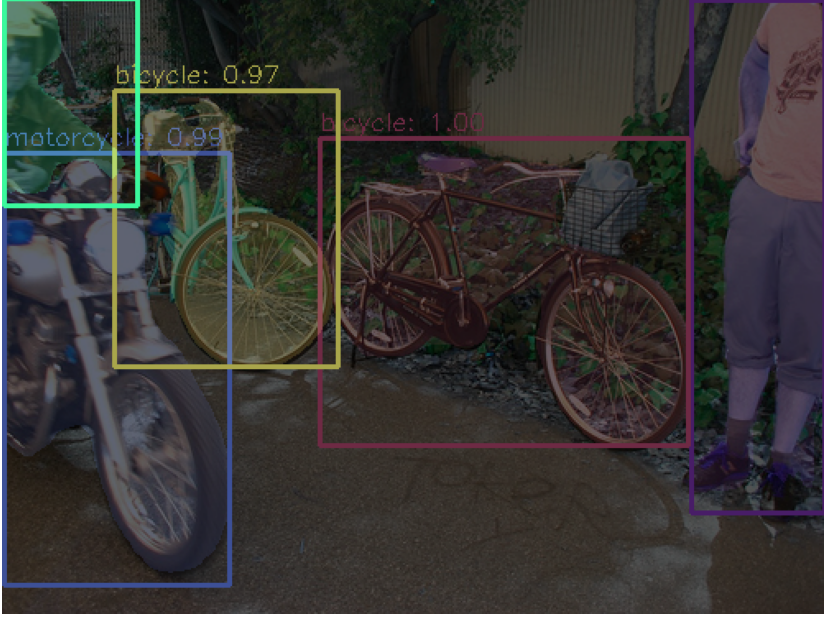}}
  \centerline{(d) AdaptICMH (0.4870)}\medskip
\end{minipage}
\hfill
\begin{minipage}[b]{0.19\linewidth}
  \centering
  \centerline{\includegraphics[width=3.5cm, height=2.7cm]{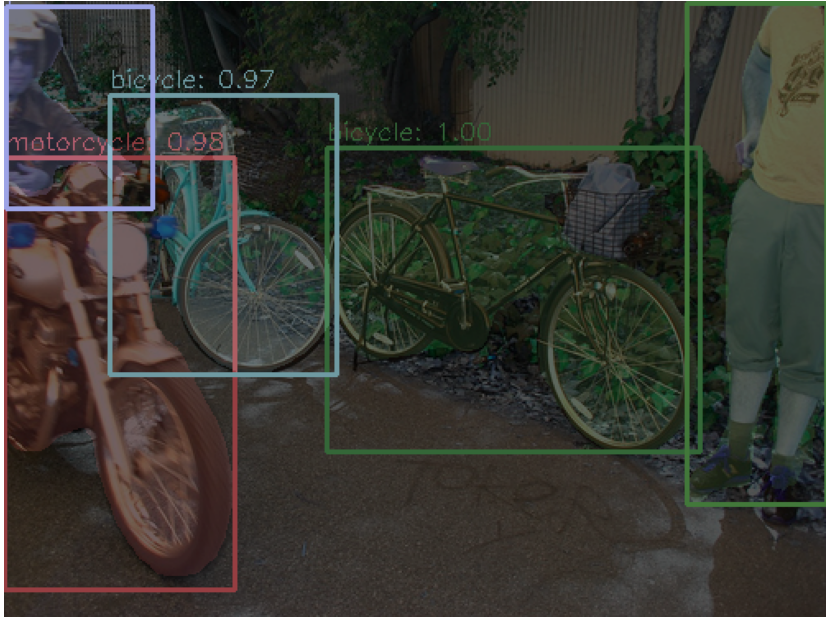}}
  \centerline{(e) CI-ICM (0.4485)}\medskip
\end{minipage}

\begin{minipage}[b]{0.19\linewidth}
  \centering
  \centerline{\includegraphics[width=3.5cm, height=4cm]{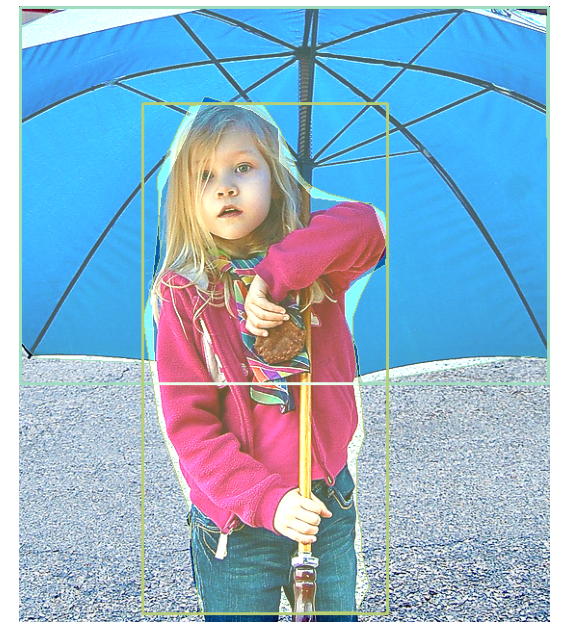}}
  \centerline{(f) Ground truth}\medskip
\end{minipage}
\hfill
\begin{minipage}[b]{0.19\linewidth}
  \centering
  \centerline{\includegraphics[width=3.5cm, height=4cm]{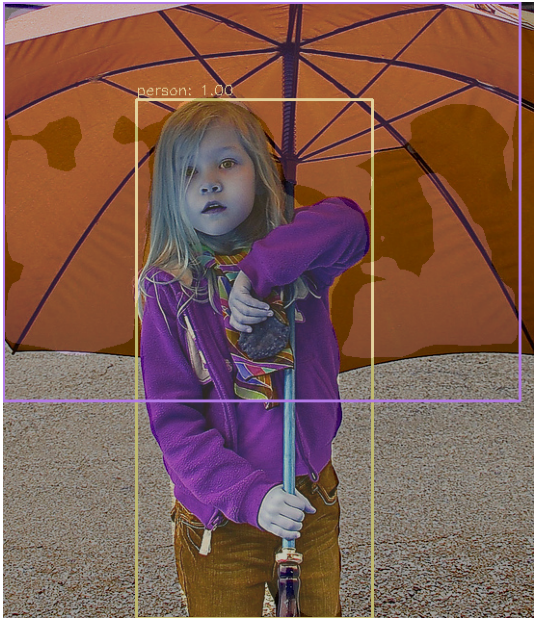}}
  \centerline{(g) ELIC (0.6608)}\medskip
\end{minipage}
\hfill
\begin{minipage}[b]{0.19\linewidth}
  \centering
  \centerline{\includegraphics[width=3.5cm, height=4cm]{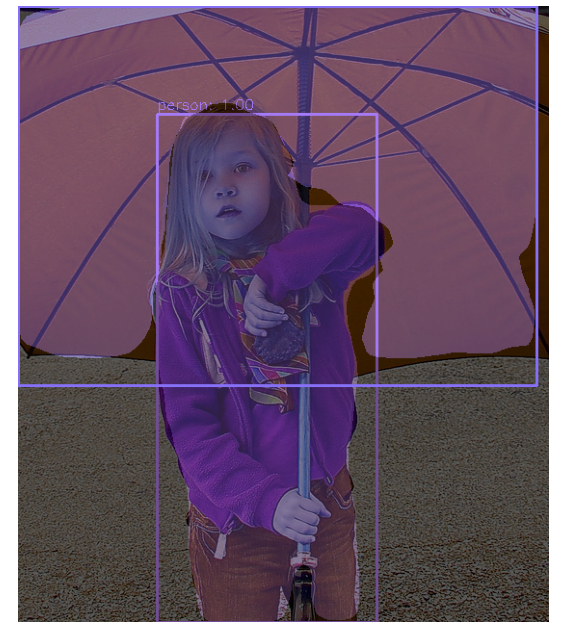}}
  \centerline{(h) TransTIC (0.4057)}\medskip
\end{minipage}
\hfill
\begin{minipage}[b]{0.19\linewidth}
  \centering
  \centerline{\includegraphics[width=3.5cm, height=4cm]{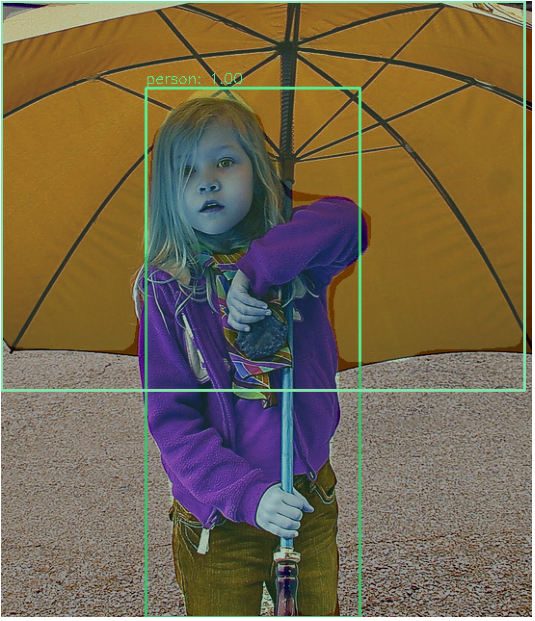}}
  \centerline{(i) AdaptICMH (0.4576)}\medskip
\end{minipage}
\hfill
\begin{minipage}[b]{0.19\linewidth}
  \centering
  \centerline{\includegraphics[width=3.5cm, height=4cm]{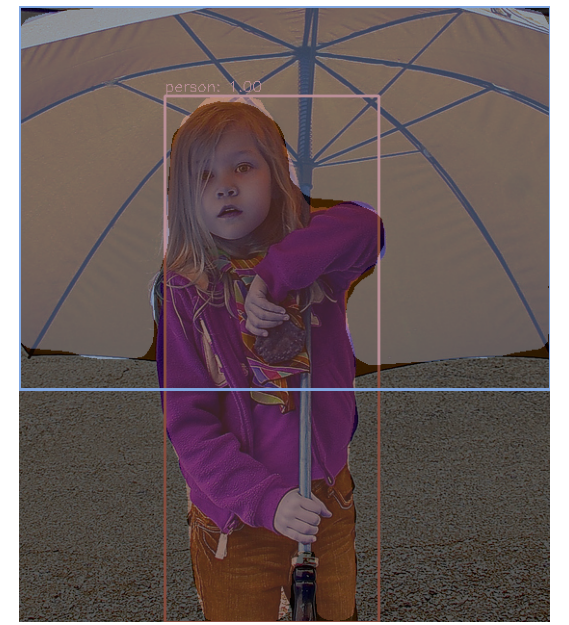}}
  \centerline{(j) CI-ICM (0.3358) }\medskip
\end{minipage}
\caption{Visualization of instance segmentation results from coded images, where bpp values also presented. (a) and (f): Ground truth, (b) and (g): ELIC, (c) and (h): TransTIC, (d) and (i): AdaptICMH, (e) and (j): the proposed CI-ICM, (a)-(e): image of person and bicycles, (f)-(j): image of a girl under an umbrella.}
\label{visualization_insseg}
\end{figure*}

\subsection{Coding Performance for Instance Segmentation}
\label{inssegexp}
In addition to object detection, the accuracy of instance segmentation on the coded images from different codecs is also evaluated. {The Mask R-CNN with ResNet-50 is employed as backbone}. Table~\ref{result table ins seg} presents quantitative BD-mAPs of instance segmentation task, comparing the proposed CI-ICM and benchmark methods against ELIC. Similar to object detection, the JPEG achieves a negative BD-mAP@50:95, i.e., -5.969$\%$, while BD-mAP@50:95s of the {VTM,} MLIC, TransTIC, AdaptICMH, and the proposed CI-ICM are {0.542$\%$,} 1.233$\%$, 10.208$\%$, 11.282$\%$ and 13.718$\%$, respectively. The proposed CI-ICM is 2.436$\%$ better than that of AdaptICMH and is the best among all benchmark schemes in terms of the BD-mAP@50:95. Similar results can be found for BD-mAP@50 and BD-mAP@75, where 3.807$\%$ and 2.653$\%$ more accuracy gains are achieved compared with the AdaptICMH.

\renewcommand{\arraystretch}{1.5}
\begin{table}
  \setlength{\tabcolsep}{4pt}
  \caption{BD-mAPs($\%$) of the proposed CI-ICM and benchmark schemes as comparing with the ELIC~\cite{He2022ELICEL} for instance segmentation task.}
  \label{result table ins seg}
  \begin{center}
  \begin{tabular}{c c c c}
    \hline
    Coding schemes & BD-mAP@50:95 & BD-mAP@50 & BD-mAP@75\\
    \hline
    ELIC\cite{He2022ELICEL} & 0 & 0 & 0\\
    JPEG\cite{skodras2001jpeg} & -5.969 & -9.150 & -6.786\\
    VTM\cite{vvc} & 0.542 & 0.165 & 0.562\\
    MLIC\cite{Jiang2022MLICME} & 1.233 & 1.654 & 1.299\\
    TransTIC\cite{Chen2023TransTICTT} & 10.208 & 13.376 & 11.535\\
    AdaptICMH\cite{Li2024ImageCF} & 11.282 & 15.221 & 12.591\\
    CI-ICM & \textbf{13.718} & \textbf{19.028} & \textbf{15.244}\\
    \hline
  \end{tabular}
  \end{center}
\end{table}
\renewcommand{\arraystretch}{1}

Fig.~\ref{fig:insseg} illustrates the rate-accuracy curves of the proposed CI-ICM and benchmark methods on instance segmentation, where the y-axis of each subfigure is mAP@50:95, mAP@50, and mAP@75, respectively. First, it can be observed that ICM schemes, including TransTIC, AdaptICMH, and the proposed CI-ICM are much better than those of the human-oriented coding schemes, i.e., JPEG, {VTM,} ELIC, and MLIC. The coding performance of the proposed CI-ICM achieves higher accuracy at the same bit rate compared with TransTIC and AdaptICMH. Third, similar trends can be found from instance segmentation task evaluated with the three accuracy metrics, which validate that the proposed CI-ICM is the best.  Fig.~\ref{visualization_insseg} presents the segmentation results of two coded images from the COCO2017 validation dataset. Comparison between (b)-(e) reveals that CI-ICM preserves more complete object masks, and comparison between (g)-(j) exhibits enhanced edge detail fidelity closer to ground truth. Moreover, the bit rates of CI-ICM are 0.4485 bpp and 0.3358 bpp, which are much lower than those of other coding schemes. Overall, the CI-ICM is efficient for instance segmentation.


\begin{figure*}[!ht]
\begin{minipage}[b]{0.24\linewidth}
  \centering
  \centerline{\includegraphics[width=\linewidth]{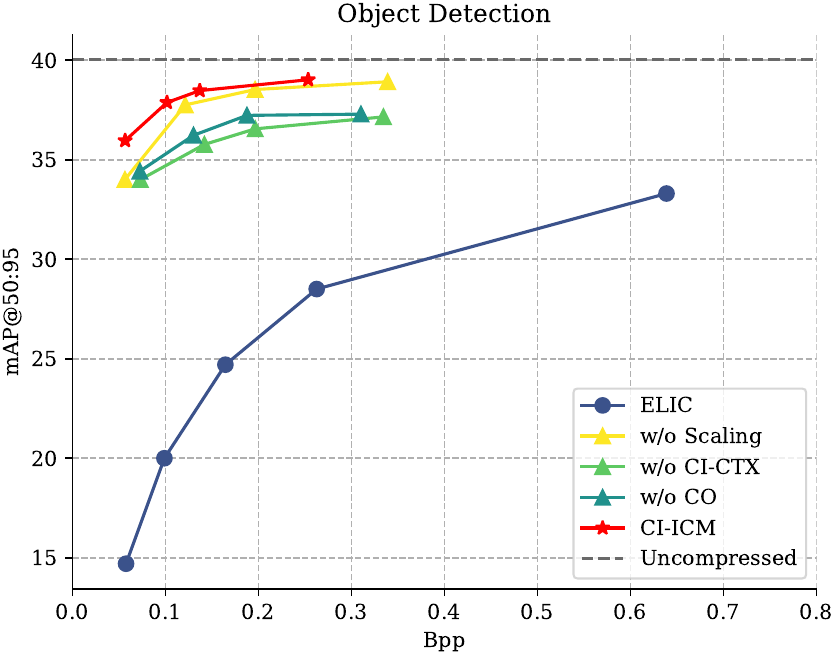}}
  \centerline{(a)}\medskip
\end{minipage}
\hfill
\begin{minipage}[b]{0.24\linewidth}
  \centering
  \centerline{\includegraphics[width=\linewidth]{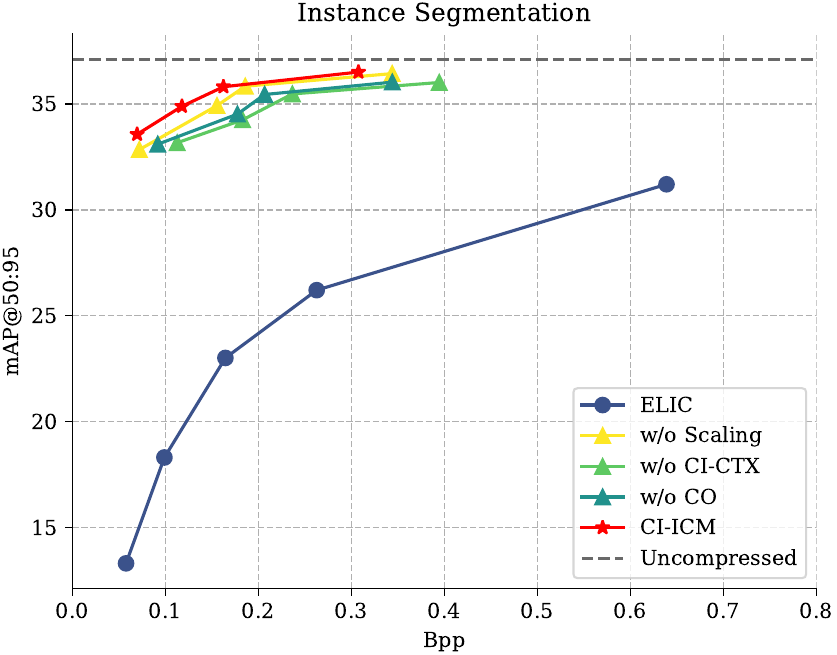}}
  \centerline{(b)}\medskip
\end{minipage}
\hfill
\begin{minipage}[b]{0.24\linewidth}
  \centering
  \centerline{\includegraphics[width=\linewidth]{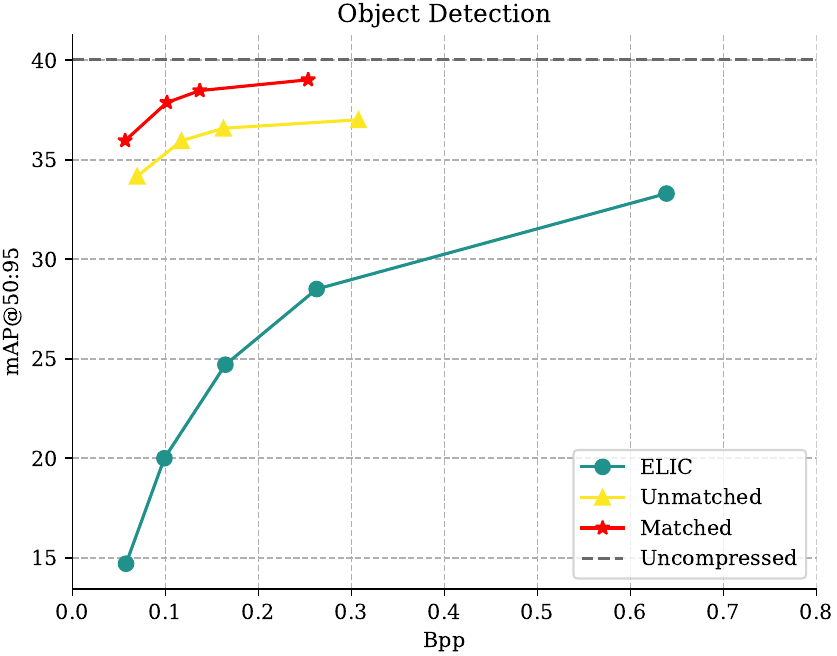}}
  \centerline{(c)}\medskip
\end{minipage}
\hfill
\begin{minipage}[b]{0.24\linewidth}
  \centering
  \centerline{\includegraphics[width=\linewidth]{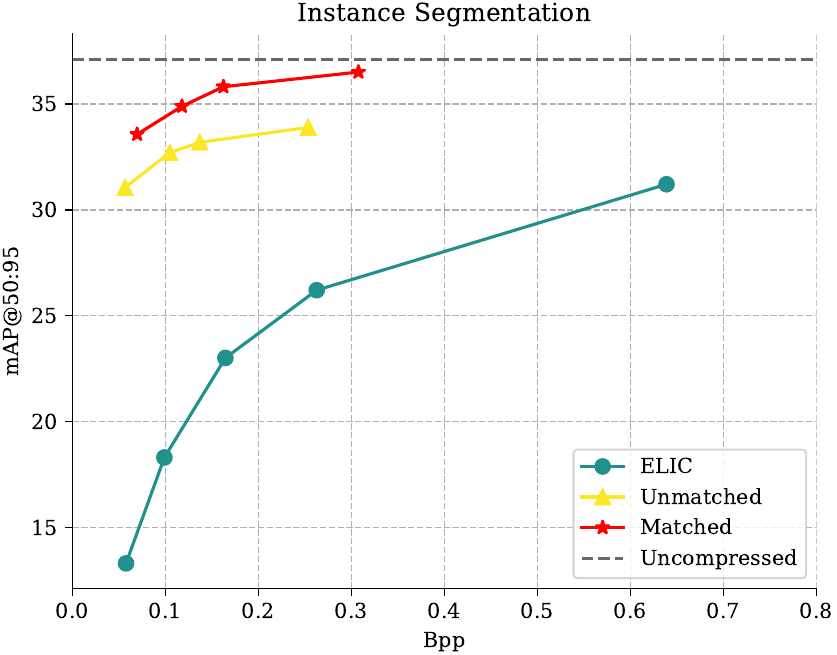}}
  \centerline{(d)}\medskip
\end{minipage}
\caption{Rate-accuracy curves of ablation studies on object detection task and instance segmentation task, where the task accuracy is measured with mAP@50:95. (a)-(b) analysis on scaling process in FCGS, CI-CTX, {CIG}, and channel order loss, (c)-(d) analysis on TSCA modes. (a)(c) object detection, and (b)(d) instance segmentation. }
\label{fig:ablstd}
\end{figure*}


It shall be noted that CIG weights, FCGS, and CI-CTX in the proposed CI-ICM are trained based on object detection task. Then, the TSCA and compression network are fine-tuned on instance segmentation. Finally, the coded images are evaluated on instance segmentation. The highest coding performance of the CI-ICM validates the importance ranking of CIG is robust, TSCA module can adapt to different tasks, and the CI-ICM framework has a good generalization ability.


\subsection{Ablation Study}
\label{ablationexp}
Ablation studies were also conducted on COCO2017 dataset to validate the effectiveness of key modules, including FCGS, CI-CTX, channel order loss, and TSCA. Object detection and instance segmentation tasks of coded images were evaluated.

\subsubsection{Ablation Study of FCGS and CI-CTX}
\label{ablcibctx}

The FCGS involves two main processes: grouping and scaling. Table~\ref{result table ablations obj} shows the BD-mAPs of the ablation studies on the scaling process in FCGS and the grouping process with CI-CTX.

We can observe that for object detection, the BD-mAP@50:95, BD-mAP@50, and BD-mAP@75 gains are 14.566$\%$, 18.834$\%$, and 16.494$\%$ if without using the scaling, labeled as ``w/o scaling", as compared with the ELIC. Compared with the full CI-ICM, it is observed that the mAP for object detection degrades 1.683$\%$, 2.090$\%$ and 1.997$\%$ in BD-mAP@50:95, BD-mAP@50, and BD-mAP@75, respectively. Similar results can be found for coded images on instance segmentation task. Fig.~\ref{fig:ablstd} shows the rate-accuracy curves of the ablation studies without using the scaling, which is inferior to the full CI-ICM. These results validate the effectiveness of scaling in FCGS in improving ICM coding performance. By adjusting the dynamic range of feature groups, the scaling process in FCGS facilitates a more effective bitrate allocation strategy, where critical task-specific feature channels are well preserved to maintain task accuracy and less important channels are quantized for a high compression ratio.

As the grouping process in FCGS is closely coupled with CI-CTX, we evaluated the coding performance of the grouping and CI-CTX together, which is labeled ``w/o CI-CTX". As shown in Table~\ref{result table ablations obj} and Fig.~\ref{fig:ablstd}, we can observe that for object detection, the BD-mAP@50:95, BD-mAP@50, and BD-mAP@75 gains are 11.838$\%$, 15.882$\%$, and 13.723$\%$, respectively, as compared to ELIC. Compared with the full CI-ICM, mAP degrades 4.411$\%$, 5.042$\%$, and 4.768$\%$ in BD-mAP@50:95, BD-mAP@50, and BD-mAP@75, respectively, which are larger than those of ``w/o scaling". Similar results can be found for instance segmentation task. These results validate the important and indispensable contribution of the grouping and CI-CTX in CI-ICM, which are able to effectively prioritize bit allocation to important feature channels.

\renewcommand{\arraystretch}{1.5}
\begin{table}
  \setlength{\tabcolsep}{1pt}
  \caption{BD-mAPs($\%$) of the ablation studies of the CI-ICM for object detection and instance segmentation, where ELIC~\cite{He2022ELICEL} is the anchor.}
  \label{result table ablations obj}
  \begin{center}
  \begin{tabular}{@{}c c c c c@{}}
    \hline
    Tasks & \makecell[c]{Coding \\ schemes} & BD-mAP@50:95 & BD-mAP@50 & BD-mAP@75\\
    \hline
    \multirow{4}{*}[0pt]{\makecell[c]{Object \\ detection}} 
    & w/o Scaling & 14.566 & 18.834 & 16.494\\
    & w/o CI-CTX & 11.838 & 15.882 & 13.723\\
    & w/o CO & 12.742 & 16.824 & 14.619\\
    & CI-ICM & \textbf{16.249} & \textbf{20.924} & \textbf{18.491}\\
    \hline
    \multirow{4}{*}[0pt]{\makecell[c]{Instance \\ segmentation}} 
    & w/o Scaling & 12.650 & 17.464 & 14.101\\
    & w/o CI-CTX & 10.251 & 13.797 & 11.545\\
    & w/o CO & 11.287 & 15.279 & 12.896\\
    & CI-ICM & \textbf{13.718} & \textbf{19.028} & \textbf{15.244}\\
    \hline
  \end{tabular}
  \end{center}
\end{table}
\renewcommand{\arraystretch}{1}

\subsubsection{Ablation Study of {CIG} and Channel Order Loss}
\label{ablcol}
In addition, we also performed an ablation study on {the CIG module and channel order loss $L_{CO}$. As the CIG is tightly coupled with channel order loss, we evaluated the coding performance of them together, which is used to extract the ordered features based on their importance to the machine vision task.} As shown in Fig.~\ref{fig:ablstd}(a)(b) and Table~\ref{result table ablations obj}. {If without the channel ordering, labeled as ``w/o CO"}, the BD-mAPs for object detection degrade 3.507$\%$, 4.100$\%$, 3.872$\%$ in mAP@50:95, BD-mAP@50 and BD-mAP@75, respectively compared with the full CI-ICM. Similarly, the BD-mAPs for the instance segmentation degrade 2.431$\%$, 3.749$\%$, and 2.348$\%$, respectively. The guidance provided by {CIG and} channel order loss ensures effective feature ordering and grouping, which is indispensable during the training process and yields compression performance gains.

\subsubsection{Ablation Study of TSCA}
\label{abltsca}


\renewcommand{\arraystretch}{1.5}
\begin{table}
  \setlength{\tabcolsep}{3pt}
  \caption{BD-mAPs($\%$) of using TSCA with ``matched" and ``unmatched" mode for object detection and instance segmentation, where ELIC~\cite{He2022ELICEL} is used as anchor.}
  \label{result table tsca}
  \begin{center}
  \begin{tabular}{@{}c c c c c@{}}
    \hline
    Tasks & Mode & BD-mAP@50:95 & BD-mAP@50 & BD-mAP@75\\
    \hline
    \multirow{2}{*}[0pt]{\makecell[c]{Object \\ detection}} & Unmatched & 12.672 & 16.938 & 14.516 \\
    & Matched & \textbf{16.249} & \textbf{20.924} & \textbf{18.491}  \\
    \hline
    \multirow{2}{*}[0pt]{\makecell[c]{Instance \\ segmentation}} & Unmatched & 12.778 & 18.646 & 13.984\\
    & Matched & \textbf{13.718} & \textbf{19.028} & \textbf{15.244}\\
    \hline
  \end{tabular}
  \end{center}
\end{table}
\renewcommand{\arraystretch}{1}

As described in Section~\ref{tsca}, the TSCA switches modes by activating different CABs for multiple tasks, which are object detection and instance segmentation in this paper. The TSCA includes two CABs and one for each task. To validate the effectiveness of the TSCA, we compare performance when using the task-matched CAB versus the task-unmatched CAB, denoted as ``Matched" and ``Unmatched", respectively. The task-matched mode indicates that the CAB is trained for task A and is tested for coding on task A with training stage 3. The task-unmatched mode is the CAB that is trained/finetuned from task A but tested for coding on task B, where the target task is unmatched with the trained task. Table~\ref{result table tsca} shows the BD-mAPs of using TSCA with matched and unmatched modes for object detection and instance segmentation. We can observe that for object detection, the BD-mAP@50:95s are 16.249 $\%$ and 12.672 $\%$ for using matched and unmatched CABs, respectively. For instance segmentation, the BD-mAP@50:95s are 13.718 $\%$ and 12.778 $\%$, for using matched and unmatched CABs, respectively. Similar results and trends are found for BD-mAP@50 and BD-mAP@75. Fig.~\ref{fig:ablstd}(c)(d) shows the rate-accuracy curves of CI-ICM using different TSCA modes, where the matched modes achieve the best. These results show that updating the TSCA with matched training Stage 3 for a specific task is effective for high efficiency coding. {In the unmatched mode, bitstreams generated for task A are directly used for task B. Compared to the matched configurations, the proposed CI-ICM still achieves significant BD-mAPs over ELIC in the unmatched mode, demonstrating the generalization ability of the ICM in cross-task applications.}

\begin{figure*}[!ht]
\begin{minipage}[b]{0.24\linewidth}
  \centering
  \centerline{\includegraphics[width=\linewidth]{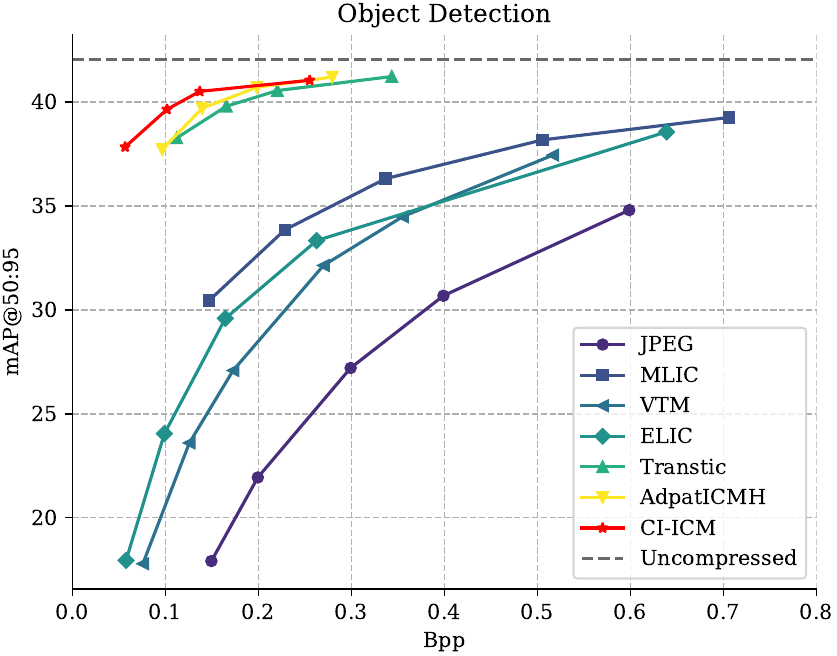}}
  \centerline{(a)}\medskip
\end{minipage}
\hfill
\begin{minipage}[b]{0.24\linewidth}
  \centering
  \centerline{\includegraphics[width=\linewidth]{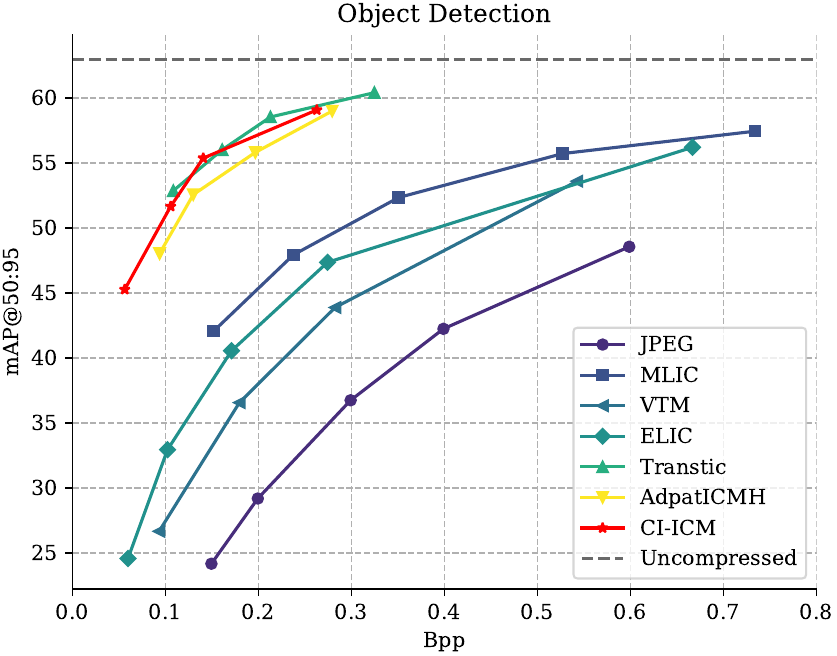}}
  \centerline{(b)}\medskip
\end{minipage}
\hfill
\begin{minipage}[b]{0.24\linewidth}
  \centering
  \centerline{\includegraphics[width=\linewidth]{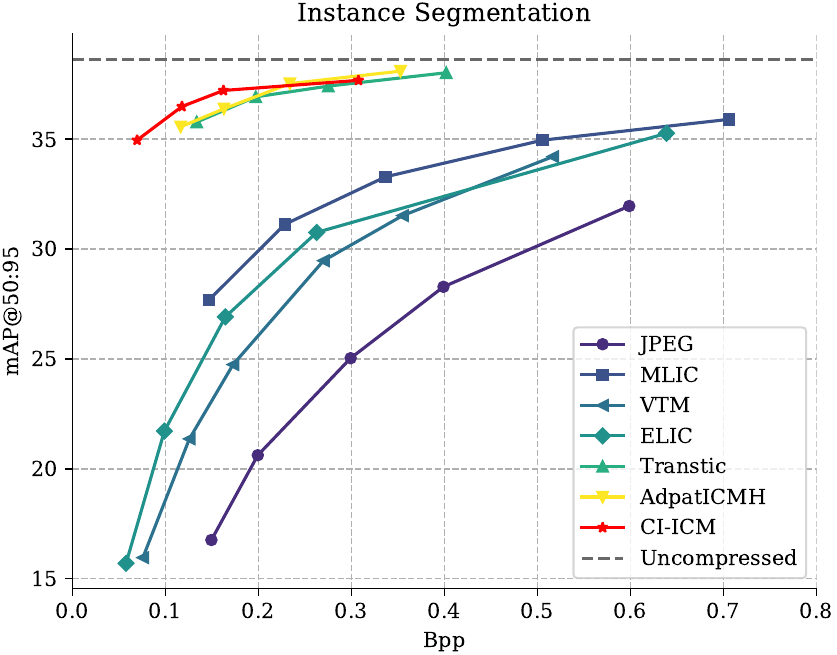}}
  \centerline{(c)}\medskip
\end{minipage}
\hfill
\begin{minipage}[b]{0.24\linewidth}
  \centering
  \centerline{\includegraphics[width=\linewidth]{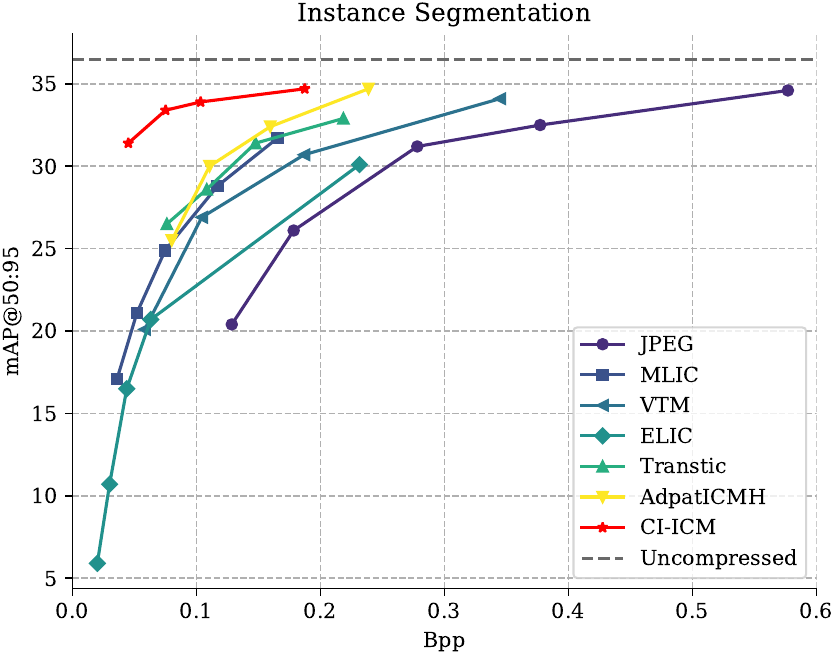}}
  \centerline{(d)}\medskip
\end{minipage}
\caption{{Rate-accuracy curves of generalization studies, where the task accuracy is measured with mAP@50:95. (a) Analysis on COCO 2017 dataset, Faster R-CNN with ResNet-101 backbone for object detection. (b) Analysis on Pascal VOC 2012 dataset, Faster R-CNN with ResNet-50 backbone for object detection. (c) Analysis on COCO 2017 dataset, Mask R-CNN with ResNet-101 backbone for instance segmentation. (d) Analysis on Cityscapes dataset, Mask R-CNN with ResNet-50 backbone for instance segmentation.}}
\label{fig:generlization}
\end{figure*}

\renewcommand{\arraystretch}{1.5}
\begin{table}
  \setlength{\tabcolsep}{4pt}
  \caption{{BD-mAP@50:95s($\%$) of the proposed CI-ICM and benchmark schemes for different datasets and feature extraction network, where ELIC~\cite{He2022ELICEL} is used as the anchor.}}
  \label{tab:generlization}
  \begin{center}
  \begin{tabular}{@{}c c c c c c@{}}
    \hline
    \multirow{2}{*}{Coding schemes} & \multicolumn{2}{c}{\makecell[c]{Object detection}} & \multicolumn{2}{c}{\makecell[c]{Instance segmentation}}\\
    & ResNet-101 & Pascal VOC & ResNet-101 & Cityscapes\\
    \hline
    ELIC\cite{He2022ELICEL} & 0 & 0 & 0 & 0\\
    JPEG\cite{skodras2001jpeg} & -7.130 & -11.267 & -6.393 & -3.934\\
    VTM\cite{vvc} & -2.190 & -3.910 & -2.076 & 0.686\\
    MLIC\cite{Jiang2022MLICME} & 1.207 & 2.112 & 1.025 & 2.239\\
    TransTIC\cite{Chen2023TransTICTT} & 9.277 & 15.294 & 7.491 & 2.507\\
    AdaptICMH\cite{Li2024ImageCF} & 10.764 & 14.379 & 8.496 & 3.164\\
    CI-ICM & \textbf{13.854} & \textbf{17.518} & \textbf{11.326} & \textbf{9.072}\\
    \hline
  \end{tabular}
  \end{center}
\end{table}
\renewcommand{\arraystretch}{1}

\subsection{{Generalization Analysis}}

{To assess the generalization capability of the CI-ICM, we conducted additional experiments across different feature extraction backbones and datasets. We used the original pretrained CI-ICM model without retraining or finetuning in all these generalization experiments.}

\subsubsection{{Generalization across Model Backbones}}
{We evaluated CI-ICM performance when using different feature extraction backbones. Specifically, we replaced ResNet-50 with ResNet-101 in Faster R-CNN and Mask R-CNN, and then tested on the COCO 2017 validation set. Results in Fig. \ref{fig:generlization}(a)(c) and Table \ref{tab:generlization} show that CI-ICM achieves BD-mAP@50:95 of 13.854$\%$ and 11.326$\%$ for object detection and instance segmentation, respectively. These results indicate that CI-ICM maintains leading performance across different task model architectures.}

\subsubsection{{Generalization across Datasets}}
{We evaluated CI-ICM on different datasets to assess cross-dataset generalization. We tested object detection on Pascal VOC 2012 dataset and instance segmentation on Cityscapes dataset, and used the original ResNet-50 backbone for both tasks. Fig. \ref{fig:generlization}(b)(d) and Table \ref{tab:generlization} show that CI-ICM achieves BD-mAP@50:95 of 17.518$\%$ and 9.072$\%$ for object detection and instance segmentation, respectively. These results indicate that CI-ICM maintains leading performance across different datasets.}

{Consistent results appear across other evaluation metrics. In summary, CI-ICM achieves robust generalization ability and competitive compression performance across different feature extraction backbones and datasets. This robustness stems from transmitting and reconstructing image features rather than task-specific outputs, which preserves critical information applicable to diverse downstream tasks and models.}

\renewcommand{\arraystretch}{1.5}
\begin{table}
  \setlength{\tabcolsep}{4pt}
  \caption{FLOPs and number of parameters comparison between the proposed CI-ICM and learning-based methods}
  \label{tab:complexity}
  \begin{center}
  \begin{tabular}{@{}c c c c c c@{}}
    \hline
    \multirow{2}{*}{Coding schemes} & \multicolumn{2}{c}{\makecell[c]{Coding time (s)}} & \multirow{2}{*}{FLOPs (G)} & \multirow{2}{*}{Params. (M)}\\
    & Encoding & Decoding & & \\
    \hline
    ELIC\cite{He2022ELICEL} & 2.722 & \textbf{0.124} & 871.420 & 31.656\\
    MLIC\cite{Jiang2022MLICME} & \textbf{2.133} & 0.203 & 1317.803 & 116.481\\
    CF\cite{end-to-end} & 2.746 & 0.251 & 180.672 & 34.755\\
    SC2\cite{matsubara2023sc} & 2.401 & 0.169 & \textbf{151.540} & 34.014\\
    TransTIC\cite{Chen2023TransTICTT} & 2.311 & 0.175 & 529.876 & 9.028\\
    AdaptICMH\cite{Li2024ImageCF} & 2.166 & 0.149 & 352.672 & \textbf{7.797}\\
    CI-ICM & 2.288 & 0.167 & 565.086 & 15.696\\
    \hline
  \end{tabular}
  \end{center}
\end{table}
\renewcommand{\arraystretch}{1}

\subsection{Computational Complexity Analysis}
\label{cca}
We also analyze the computational complexity of the proposed CI-ICM. Learning based coding schemes are used as benchmarks for comparison.
The coding experiments were carried out on a server equipped with NVIDIA GeForce RTX 3090 GPU, and the resolution of coded images is $1024\times1024$ from COCO2017. Table~\ref{tab:complexity} shows complexity analysis of the proposed CI-ICM in terms of coding times, FLOPs, and the number of model parameters. First, MLIC has the highest model size and computing FLOPs, and SC2 has the lowest computational cost, but is not the worst-performing method. This decoupling suggests that the model complexity and ICM coding performance are not directly correlated. Effective design and optimization are critical. Second, the encoding time of the proposed CI-ICM is 2.288s, which is lower than that of TransTIC, SC2, CF, and ELIC. The decoding time is 0.167s, which is much lower than the encoding time. Moreover, it is lower than MLIC, CF, SC2, TransTIC, but is higher than AdaptICMH and ELIC. As for the FLOPs, the proposed CI-ICM is 565.086G and is similar to TransTIC. The number of the model parameters is 15.696M, which is much lower than ELIC, CF, SC2, and MLIC, but a little higher than TransTIC and AdaptICMH. The CI-ICM achieves higher coding efficiency with moderate computational complexity and model size. 

\section{Conclusions}
\label{conclusion}
In this paper, we propose a Channel Importance-driven Learned Image Coding for Machines (CI-ICM), which improves the coding efficiency by exploiting the varying importance of feature channels. First, we propose a Channel Importance Generation (CIG) module to analyze the importance of feature channels. Then, we propose Feature Channel Grouping and Scaling (FCGS) and Channel Importance-based Context (CI-CTX) modules to allocate more bits to feature channels with higher-importance and less bits to lower-importance channels for a high compression ratio. Third, we propose a Task-Specific Channel Adaptation (TSCA) module to adapt to multiple different machine vision tasks and enhance the applicability. Experimental results, ablation studies, and complexity analysis validated the effectiveness, complexity, and applicability of the proposed CI-ICM and key modules.

\ifCLASSOPTIONcaptionsoff
  \newpage
\fi

\bibliographystyle{IEEEtran}
\bibliography{reference}

\begin{thebibliography}{10}
\providecommand{\url}[1]{#1}
\csname url@samestyle\endcsname
\providecommand{\newblock}{\relax}
\providecommand{\bibinfo}[2]{#2}
\providecommand{\BIBentrySTDinterwordspacing}{\spaceskip=0pt\relax}
\providecommand{\BIBentryALTinterwordstretchfactor}{4}
\providecommand{\BIBentryALTinterwordspacing}{\spaceskip=\fontdimen2\font plus
\BIBentryALTinterwordstretchfactor\fontdimen3\font minus \fontdimen4\font\relax}
\providecommand{\BIBforeignlanguage}[2]{{%
\expandafter\ifx\csname l@#1\endcsname\relax
\typeout{** WARNING: IEEEtran.bst: No hyphenation pattern has been}%
\typeout{** loaded for the language `#1'. Using the pattern for}%
\typeout{** the default language instead.}%
\else
\language=\csname l@#1\endcsname
\fi
#2}}
\providecommand{\BIBdecl}{\relax}
\BIBdecl

\bibitem{digitalretina}
Y.~Lou, L.-Y. Duan, Y.~Luo, Z.~Chen, T.~Liu, S.~Wang, and W.~Gao, ``Towards efficient front-end visual sensing for digital retina: A model-centric paradigm,'' \emph{IEEE Trans. Multimedia}, vol.~22, no.~11, pp. 3002--3013, Nov. 2020.

\bibitem{iot}
J.~Zhang and D.~Tao, ``Empowering things with intelligence: A survey of the progress, challenges, and opportunities in artificial intelligence of things,'' \emph{IEEE Internet Things J.}, vol.~8, no.~10, pp. 7789--7817, May 2021.

\bibitem{iot2}
P.~Zhang, F.~Huang, D.~Wu, B.~Yang, Z.~Yang, and L.~Tan, ``Device-edge-cloud collaborative acceleration method towards occluded face recognition in high-traffic areas,'' \emph{IEEE Trans. Multimedia}, vol.~25, pp. 1513--1520, Mar. 2023.

\bibitem{vvc}
B.~Bross, Y.-K. Wang, Y.~Ye, S.~Liu, J.~Chen, G.~J. Sullivan, and J.-R. Ohm, ``Overview of the versatile video coding (vvc) standard and its applications,'' \emph{IEEE Trans. Circuit Syst. Video Technol.}, vol.~31, no.~10, pp. 3736--3764, Oct. 2021.

\bibitem{HNRISC}
P.~Zhang, S.~Wang, M.~Wang, P.~Chen, W.~Wu, X.~Wang, and S.~Kwong, ``Hnr-isc: Hybrid neural representation for image set compression,'' \emph{IEEE Trans. Multimedia}, vol.~27, pp. 28--40, Dec. 2025.

\bibitem{vcm}
W.-H. Peng and H.-M. Hang, ``Recent advances in end-to-end learned image and video compression,'' in \emph{IEEE Int. Conf. Vis. Commun. Image Process.}, Macau, China, Dec. 2020, pp. 1--2.

\bibitem{vcmaparadigm}
L.~Duan, J.~Liu, W.~Yang, T.~Huang, and W.~Gao, ``Video coding for machines: A paradigm of collaborative compression and intelligent analytics,'' \emph{IEEE Trans. Image Process.}, vol.~29, pp. 8680--8695, Aug. 2020.

\bibitem{TDVCM}
X.~Yi, H.~Wang, S.~Kwong, and C.-C. Jay~Kuo, ``Task-driven video compression for humans and machines: Framework design and optimization,'' \emph{IEEE Trans. Multimedia}, vol.~25, pp. 8091--8102, Dec. 2023.

\bibitem{JinJND2022}
J.~Jin, X.~Zhang, X.~Fu, H.~Zhang, W.~Lin, J.~Lou, and Y.~Zhao, ``Just noticeable difference for deep machine vision,'' \emph{IEEE Trans. Circuit Syst. Video Technol.}, vol.~32, no.~6, pp. 3452--3461, Jun. 2022.

\bibitem{TAQNforJPEG}
J.~Choi and B.~Han, ``Task-aware quantization network for jpeg image compression,'' in \emph{Eur. Conf. Comput. Vis.}, Nov. 2020, pp. 309--324.

\bibitem{analysisRD2021}
Z.~Huang, C.~Jia, S.~Wang, and S.~Ma, ``Visual analysis motivated rate-distortion model for image coding,'' in \emph{IEEE Int. Conf. Multimedia Expo}, Shenzhen, China, Jun. 2021, pp. 1--6.

\bibitem{SaliencyBA}
Y.~Li, W.~Gao, G.~Li, and S.~Ma, ``Saliency segmentation oriented deep image compression with novel bit allocation,'' \emph{IEEE Trans. Image Process.}, vol.~34, pp. 16--29, Nov. 2025.

\bibitem{zhangyun2024}
Y.~Zhang, H.~Lin, J.~Sun, L.~Zhu, and S.~Kwong, ``Learning to predict object-wise just recognizable distortion for image and video compression,'' \emph{IEEE Trans. Multimedia}, vol.~26, pp. 5925--5938, Dec. 2024.

\bibitem{towardsSIC2021}
S.~Yang, Y.~Hu, W.~Yang, L.-Y. Duan, and J.~Liu, ``Towards coding for human and machine vision: Scalable face image coding,'' \emph{IEEE Trans. Multimedia}, vol.~23, pp. 2957--2971, Mar. 2021.

\bibitem{shindo2024image}
T.~Shindo, K.~Yamada, T.~Watanabe, and H.~Watanabe, ``Image coding for machines with edge information learning using segment anything,'' in \emph{IEEE Int. Conf. Image Process.}, Abu Dhabi, UAE, Oct. 2024, pp. 3702--3708.

\bibitem{zhang2023rethinking}
P.~Zhang, S.~Wang, M.~Wang, J.~Li, X.~Wang, and S.~Kwong, ``Rethinking semantic image compression: Scalable representation with cross-modality transfer,'' \emph{IEEE Trans. Circuit Syst. Video Technol.}, vol.~33, no.~8, pp. 4441--4445, Aug. 2023.

\bibitem{endtoend2021wang}
S.~Wang, Z.~Wang, S.~Wang, and Y.~Ye, ``End-to-end compression towards machine vision: Network architecture design and optimization,'' \emph{IEEE Open J. Circuits Syst.}, vol.~2, pp. 675--685, Nov. 2021.

\bibitem{ICMcontent2021}
N.~Le, H.~Zhang, F.~Cricri, R.~Ghaznavi-Youvalari, H.~R. Tavakoli, and E.~Rahtu, ``Learned image coding for machines: A content-adaptive approach,'' in \emph{IEEE Int. Conf. Multimedia Expo}, Shenzhen, China, Nov. 2021, pp. 1--6.

\bibitem{LeICM2021}
N.~Le, H.~Zhang, F.~Cricri, R.~Ghaznavi-Youvalari, and E.~Rahtu, ``Image coding for machines: an end-to-end learned approach,'' in \emph{ICASSP IEEE Int. Conf. Acoust. Speech Signal Process}, Jun. 2021, pp. 1590--1594.

\bibitem{end-to-end}
S.~Singh, S.~Abu-El-Haija, N.~Johnston, J.~Ballé, A.~Shrivastava, and G.~Toderici, ``End-to-end learning of compressible features,'' in \emph{IEEE Int. Conf. Image Process.}, Abu Dhabi, UAE, Oct. 2020, pp. 3349--3353.

\bibitem{matsubara2023sc}
Y.~Matsubara, R.~Yang, M.~Levorato, and S.~Mandt, ``{SC}2 benchmark: Supervised compression for split computing,'' \emph{Trans. Mach. Learn. Res.}, pp. 1--20, Jun. 2023.

\bibitem{liu2025multiscale}
J.~Liu, Y.~Zhang, Z.~Guo, X.~Huang, and G.~Jiang, ``Multiscale feature importance-based bit allocation for end-to-end feature coding for machines,'' \emph{ACM Trans. Multimed. Comput. Commun. Appl.}, vol.~21, no.~9, pp. 1--19, Sep. 2025.

\bibitem{endtoendICM2021}
L.~D. Chamain, F.~Racapé, J.~Bégaint, A.~Pushparaja, and S.~Feltman, ``End-to-end optimized image compression for machines, a study,'' in \emph{Data Compression Conf.}, Snowbird, UT, USA, Mar. 2021, pp. 163--172.

\bibitem{Rate-DistortionTheory2025}
A.~Harell, Y.~Foroutan, N.~Ahuja, P.~Datta, B.~Kanzariya, V.~S. Somayazulu, O.~Tickoo, A.~de~Andrade, and I.~V. Bajić, ``Rate-distortion theory in coding for machines and its applications,'' \emph{IEEE Trans. Pattern Anal. Mach. Intell.}, vol.~47, no.~7, pp. 5501--5519, Jul. 2025.

\bibitem{improvingmvtincompressdomain}
J.~Liu, H.~Sun, and J.~Katto, ``Improving multiple machine vision tasks in the compressed domain,'' in \emph{Int. Conf. Pattern Recog.}, Montreal, QC, Canada, Aug. 2022, pp. 331--337.

\bibitem{scalableicm_choi}
H.~Choi and I.~V. Bajić, ``Scalable image coding for humans and machines,'' \emph{IEEE Trans. Image Process.}, vol.~31, pp. 2739--2754, Mar. 2022.

\bibitem{Choi2021LatentSpaceSF}
H.~Choi\mbox{} and I.~V. Bajić, ``Latent-space scalability for multi-task collaborative intelligence,'' in \emph{IEEE Int. Conf. Image Process.}, Anchorage, AK, USA, Sep. 2021, pp. 3562--3566.

\bibitem{zhang2024}
G.~Zhang, X.~Zhang, and L.~Tang, ``Unified and scalable deep image compression framework for human and machine,'' \emph{ACM Trans. Multimedia Comput. Commun. Appl.}, vol.~20, no.~10, pp. 1--22, Oct. 2024.

\bibitem{latentSICM}
E.~Özyılkan, M.~Ulhaq, H.~Choi, and F.~Racapé, ``Learned disentangled latent representations for scalable image coding for humans and machines,'' in \emph{Data Compression Conf.}, Snowbird, UT, USA, Mar. 2023, pp. 42--51.

\bibitem{mutualFCMV2023}
T.~Liu, M.~Xu, S.~Li, C.~Chen, L.~Yang, and Z.~Lv, ``Learnt mutual feature compression for machine vision,'' in \emph{IEEE Int. Conf. Acoust. Speech Signal Process.}, Rhodes Island, Greece, Jun. 2023, pp. 1--5.

\bibitem{SSICMfeature}
N.~Yan, D.~Liu, H.~Li, and F.~Wu, ``Semantically scalable image coding with compression of feature maps,'' in \emph{IEEE Int. Conf. Image Process.}, Abu Dhabi, UAE, Oct. 2020, pp. 3114--3118.

\bibitem{Cui2025}
T.~Cui, Y.~Wang, Y.~Wang, and Z.~Fang, ``Semantic and saliency-aware scalable image coding towards human-machine collaboration,'' \emph{IEEE Trans. Circuit Syst. Video Technol.}, pp. 1--1, May 2025.

\bibitem{Chen2023TransTICTT}
Y.-H. Chen, Y.~Weng, C.-H. Kao, C.~Chien, W.-C. Chiu, and W.~Peng, ``Transtic: Transferring transformer-based image compression from human perception to machine perception,'' in \emph{IEEE/CVF Int. Conf. Comput. Vis.}, Paris, France, Oct. 2023, pp. 23\,240--23\,250.

\bibitem{Li2024ImageCF}
H.~Li, S.~Li, S.~Ding, W.~Dai, M.~Cao, C.~Li, J.~Zou, and H.~Xiong, ``Image compression for machine and human vision with spatial-frequency adaptation,'' in \emph{Eur. Conf. Comput. Vis.}, Milan, Italy, Oct. 2024, pp. 382--399.

\bibitem{fasterrcnn}
S.~Ren, K.~He, R.~Girshick, and J.~Sun, ``Faster r-cnn: Towards real-time object detection with region proposal networks,'' \emph{IEEE Trans. Pattern Anal. Mach. Intell.}, vol.~39, no.~6, pp. 1137--1149, Jun. 2017.

\bibitem{He2022ELICEL}
D.~He, Z.~Yang, W.~Peng, R.~Ma, H.~Qin, and Y.~Wang, ``Elic: Efficient learned image compression with unevenly grouped space-channel contextual adaptive coding,'' in \emph{IEEE/CVF Conf. Comput. Vis. Pattern Recog.}, New Orleans, LA, USA, Jun. 2022, pp. 5708--5717.

\bibitem{senet}
J.~Hu, L.~Shen, and G.~Sun, ``Squeeze-and-excitation networks,'' in \emph{IEEE/CVF Conf. Comput. Vis. Pattern Recog.}, Salt Lake City, UT, USA, Jun. 2018, pp. 7132--7141.

\bibitem{Jiang2022MLICME}
W.~Jiang, J.~Yang, Y.~Zhai, P.~Ning, F.~Gao, and R.~Wang, ``Mlic: Multi-reference entropy model for learned image compression,'' in \emph{ACM Int. Conf. Multimedia}, New York, NY, USA, Oct. 2023, pp. 7618 -- 7627.

\bibitem{GOSWAMI2024111921}
P.~Goswami, L.~Aggarwal, A.~Kumar, R.~Kanwar, and U.~Vasisht, ``Real-time evaluation of object detection models across open world scenarios,'' \emph{Appl. Soft Comput.}, vol. 163, p. 111921, Sep. 2024.

\bibitem{monotonicity_loss}
A.~R. Gonzales, C.~Amrhein, N.~Aepli, and R.~Sennrich, ``On biasing transformer attention towards monotonicity,'' in \emph{N. Am. Chapter Assoc. Comput. Linguist.}\hskip 1em plus 0.5em minus 0.4em\relax Online: Association for Computational Linguistics, Jun. 2021, pp. 4474--4488.

\bibitem{He2017MaskR}
K.~He, G.~Gkioxari, P.~Doll{\'a}r, and R.~Girshick, ``Mask r-cnn,'' in \emph{IEEE Int. Conf. Comput. Vis.}, Venice, Italy, Oct. 2017, pp. 2980--2988.

\bibitem{skodras2001jpeg}
A.~Skodras, C.~Christopoulos, and T.~Ebrahimi, ``The jpeg 2000 still image compression standard,'' \emph{IEEE Signal Process. Mag.}, vol.~18, no.~5, pp. 36--58, Sep. 2001.

\bibitem{miou}
M.~Everingham, L.~Van~Gool, C.~K. Williams, J.~Winn, and A.~Zisserman, ``The pascal visual object classes (voc) challenge,'' \emph{Int. J. Comput. Vision}, vol.~88, no.~2, pp. 303--338, Sep. 2010.

\end{thebibliography}

\begin{IEEEbiography}[{\includegraphics[width=1in,height=1.25in,clip,keepaspectratio]{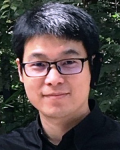}}]{Yun Zhang}
(Senior Member, IEEE) received the B.S. and M.S. degrees in electrical engineering from Ningbo University, Ningbo, China, in 2004 and 2007, respectively, and the Ph.D. degree in computer science from the Institute of Computing Technology, Chinese Academy of Sciences (CAS), Beijing, China, in 2010. From 2009 to 2014, he was a Visiting Scholar with the Department of Computer Science, City University of Hong Kong, Kowloon, Hong Kong. From 2010 to 2022, he was a Professor/Associate Professor with the Shenzhen Institutes of Advanced Technology, CAS, Shenzhen, China. Since 2022, he has been a Professor with the School of Electronics and Communication Engineering, Sun Yat-sen University (Shenzhen Campus), Guangdong, China. His research interests mainly include 3D visual signal processing, image/video compression, visual perception, and machine learning.
\end{IEEEbiography}

\vspace{-1.5\baselineskip}
\begin{IEEEbiography}[{\includegraphics[width=1in,height=1.25in,clip,keepaspectratio]{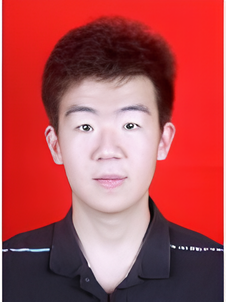}}]{Junle Liu}
received the B.E. degree in electronic information science and technology from the Sun Yat-sen University, China, in 2023. He is currently pursuing the M.E. degree with the School of Electronics and Communication Engineering, Sun Yat-sen University, Shenzhen, China. His current research interests include image coding for machines, machine vision, and deep learning.
\end{IEEEbiography}

\vspace{-1.5\baselineskip}
\begin{IEEEbiography}[{\includegraphics[width=1in,height=1.25in,clip,keepaspectratio]{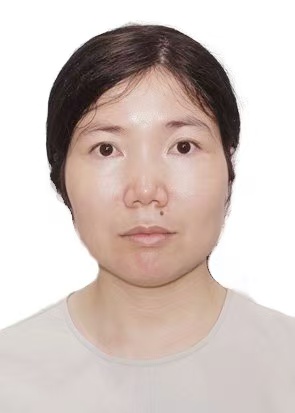}}]{Huan Zhang}
received the B.S. degree from Civil Avia tion University of China, Tianjin, China, in 2010, M.S. degree from Tsinghua University, Beijing, China in 2013, and Ph.D. degree from University of Chinese Academy of Sciences in 2021. She is currently with the School of Information Engineering, Guangdong University of Technology, Guangzhou, China.  Her research interests include image restoration, 3D image/video quality assessment, and learned image compression.
\end{IEEEbiography}

\vspace{-1.5\baselineskip}
\begin{IEEEbiography}[{\includegraphics[width=1in,height=1.25in,clip,keepaspectratio]{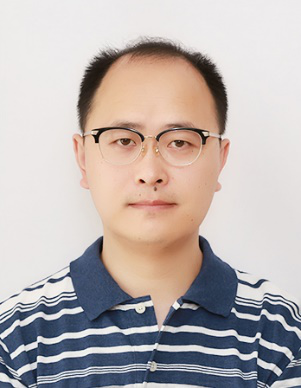}}]{Zhaoqing Pan}
(Senior Member, IEEE) received the Ph.D. degree in computer science from the City University of Hong Kong, Hong Kong, SAR, China, in 2014. In 2013, he was a Visiting Scholar with the Department of Electrical Engineering, University of Washington, Seattle, WA, USA, for six months. He is currently a Professor with the School of Electrical and Information Engineering, Tianjin University, Tianjin, China. His research interests include video coding, image quality assessment, and machine learning. He is also an Editorial Board Member of Computers and Education: X Reality and PLOS One.
\end{IEEEbiography}

\vspace{-1.5\baselineskip}
\begin{IEEEbiography}[{\includegraphics[width=1in,height=1.25in,clip,keepaspectratio]{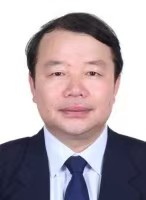}}]{Gangyi Jiang}
(Senior Member, IEEE) received the M.S. degree from Hangzhou University in 1992, and the Ph.D. degree from Ajou University, South Korea, in 2000. He is currently a Professor with the Faculty of Information Science and Engineering, Ningbo University, China. His research interests mainly include digital video compression and communications, multi-view video coding and image processing.
\end{IEEEbiography}

\vspace{-1.5\baselineskip}
\begin{IEEEbiography}[{\includegraphics[width=1in,height=1.25in,clip,keepaspectratio]{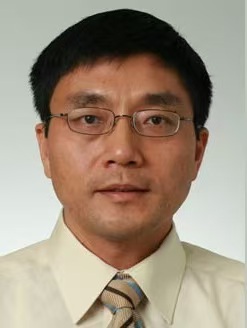}}]{Weisi Lin}
(Fellow, IEEE) received the Ph.D. degree from King’s College London, U.K. He was the Laboratory Head of visual processing with the Institute for Infocomm Research, Singapore. He is currently a Professor with the School of Computer Engineering, Nanyang Technological University, Singapore. His research interests include image processing, perceptual signal modeling, video compression, and multimedia communication, in which he has published more than 200 journal articles, more than 230 conference papers, filed 11 patents, and authored two books. He is a fellow of IET and an Honorary Fellow of the Singapore Institute of Engineering Technologists. He has been the Technical Program Chair of the IEEE ICME in 2013, the PCM in 2012, the QoMEX in 2014, and the IEEE VCIP in 2017. He has been an Associate Editor of the IEEE TRANSACTIONS ON IMAGE PROCESSING, the IEEE TRANSACTIONS ON CIRCUITS AND SYSTEMS FOR VIDEO TECHNOLOGY, the IEEE TRANSACTIONS ON MULTIMEDIA, the IEEE SIGNAL PROCESSING LETTERS, and the Journal of Visual Communication and Image Representation. He was a Distinguished Lecturer of the IEEE Circuits and Systems Society from 2016 to 2017 and the Asia–Pacific Signal and Information Processing Association (APSIPA) from 2012 to 2013. He has been an invited/panelist/keynote/tutorial speaker for more than 20 international conferences.
\end{IEEEbiography}

\end{document}